\documentclass{article}
\usepackage{PRIMEarxiv}
\usepackage{amsfonts}       
\usepackage{nicefrac}       
\usepackage{microtype}      
\usepackage{fancyhdr}       
\usepackage[hidelinks]{hyperref}

\usepackage{graphicx}
\usepackage[utf8]{inputenc}
\usepackage[english]{babel}
\usepackage{bm}
\usepackage{booktabs}
\usepackage{multirow}
\usepackage{mathtools}
\usepackage{xcolor}
\usepackage[symbol]{footmisc}

\usepackage[T1, OT1]{fontenc}
\DeclareTextSymbolDefault{\DH}{T1}
\usepackage{siunitx}
\usepackage{physics}
\usepackage{cleveref}
\usepackage{natbib}
\DeclareMathOperator*{\argmax}{arg\,max}

\newcommand{\tscale}{\tau}

\title{Emergence of Chemotactic Strategies with Multi-Agent Reinforcement Learning}

\author{
  Samuel Tovey\footnotemark[2], Christoph Lohrmann\footnotemark[2], Christian Holm \\
  Institute for Computational Physics \\
  University of Stuttgart \\
  70569, Stuttgart, Germany\\
  \texttt{\{stovey, clohrmann, holm\}@icp.uni-stuttgart.de} \\
}

\begin{document}
\maketitle
\footnotetext[2]{These authors contributed equally}

\begin{abstract}
 Reinforcement learning (RL) is a flexible and efficient method for programming micro-robots in complex environments.
Here we investigate whether reinforcement learning can provide insights into biological systems when trained to perform chemotaxis.
Namely, whether we can learn about how intelligent agents process given information in order to swim towards a target.
We run simulations covering a range of agent shapes, sizes, and swim speeds to determine if the physical constraints on biological swimmers, namely Brownian motion, lead to regions where reinforcement learners' training fails.
We find that the RL agents can perform chemotaxis as soon as it is physically possible and, in some cases, even before the active swimming overpowers the stochastic environment.
We study the efficiency of the emergent policy and identify convergence in agent size and swim speeds.
Finally, we study the strategy adopted by the reinforcement learning algorithm to explain how the agents perform their tasks.
To this end, we identify three emerging dominant strategies and several rare approaches taken.
These strategies, whilst producing almost identical trajectories in simulation, are distinct and give insight into the possible mechanisms behind which biological agents explore their environment and respond to changing conditions.

\end{abstract}

\section{Introduction}
\label{sec:introduction}
Microswimmers have the unique privilege of having evolved over millions of years to learn how to optimally navigate noisy, Brownian motion-dominated environments in search of better living conditions.
Most interactions humans are familiar with occur on length and time scales that are not subject to this noise. 
Therefore, we do naturally have an understanding of how these microswimmers can perform this navigation. 
However, understanding the emergence of this behaviour is critical as scientists strive to construct the artificial counterparts of biological microswimmers.
Previous reviews have discussed the emergence and function of biological microswimmers in great detail~\citep{bastos18a, elgeti15a}, elucidating the mechanisms and strategies behind their movement.
One recurring form of navigation in microswimmers is the so-called \texttt{run-and-tumble} motion exhibited by Escherichia coli (E-coli) wherein the bacteria will travel in a straight line for some extended period before spontaneously rotating into a random new direction~\citep{watari10a, berg04a, darnton06a}.
One application of this navigation mechanism is bacterial chemotaxis~\citep{hansen07a}, the biased movement of a bacteria towards regions with higher concentrations of beneficial chemicals or lower concentrations of harmful chemicals~\citep{wadhams04a}.
Bacteria achieve this biased motion through changes in run duration depending on changes in the concentration of the chemo-attractant or -repellant.
Learning chemotaxis has been the focal point of several research papers aimed at reproducing or better understanding biological microswimmers through the use of reinforcement learning (RL)~\citep{tovey23c, mo22a, hartl21a, landin21a}.
In their 2021 study,~\citet{hartl21a} applied a genetic algorithm to the problem of learning shape deformations for navigation in static and dynamic environments.
They found that the neural networks learned a movement closely resembling that of \texttt{run-and-tumble} motion.
In another 2021 study,~\citet{landin21a} applied Q-learning to learning navigation strategies in self-thermophoretic particles from which they again see the emergence of \texttt{run-and-tumble} motion.
They further investigated the effects of temperature on the learning process, identifying that models trained at higher temperatures took longer to learn their emergent strategy.
Finally, our previous work~\citep{tovey23c} directly addressed the role of temperature in the emergent strategy of RL-driven microswimmers by studying chemotaxis learning by the actor-critic reinforcement learning algorithm.
It was found that, while the efficacy of the chemotaxis changed with different temperatures, the same \texttt{run-and-tumble} motion arose from the majority of agents trained at different temperatures.
While it is clear that RL algorithms can and, in fact, seemingly often do learn \texttt{run-and-tumble} type motion for chemotaxis problems, what impact this has on our understanding of biological microswimmers and even optimal design of artificial swimmers is not clear.
This study investigates natural limitations on emergent chemotaxis by training actor-critic RL models using prolate, oblate, and spherical agents of different sizes and with different swim speeds in physically realistic fluid environments subject to translational and rotational Brownian motion.
In this way, we hope to identify how optimal RL algorithms are for the learning task and to identify, if any exist, optimal size/speed combinations of microswimmers in these environments, which may guide our interpretation of biological microswimmers as well as advise the design of artificial ones.
Furthermore, by investigating the deployment of the RL algorithms close to conditions where agents will be dominated by rotational and translational Brownian motion, we can explore the emergence of different navigation strategies that may be leveraged in the treatment of biological or artificial swimmers, essentially peering into the minds of bacteria as they navigate environments.

The manuscript is structured as follows. 
We will first discuss the theory behind the investigation, explaining the mechanism and physical limitations of chemotaxis in biological systems.
We then introduce deep actor-critic reinforcement learning and discuss its multi-agent realization.
The simulation and training methods are discussed in detail before our results are presented, and a brief outlook is presented.

\section{Theory}
\label{sec:theory}
\subsection{Biological Chemotaxis}
\label{sec:bio-chemo}
As mentioned briefly in the introduction, chemotaxis is the biased movement of bacteria towards favourable regions in their environment~\citep{wadhams04a}.
\textit{Escherichia coli} (e. coli) perform chemotaxis actively using \texttt{run-and-tumble} motion.
In their running phase, they bundle their flagella together and rotate them anti-clockwise, during a tumble phase, one or more flagella change their rotation direction, breaking apart the bundle and causing random rotation of the bacteria before the bundle reforms and translational swimming is resumed~\citep{turner00a}.
Utilizing a sensing mechanism, these bacteria can identify if they are moving towards or away from favourable regions of their environment, adjusting their tumble rates accordingly to maintain desired movement~\citep{schnitzer90a}.

In order to capture the essential features of bacterial motility, we consider here active particles that can perform four distinct types of movement.
Firstly, they can move forward along their intrinsic direction like bacteria in the run phase.
Secondly, they can actively rotate clockwise or counterclockwise, like bacteria in the tumble phase. 
Our swimmers can choose the direction of rotation while bacteria rotate towards a random new orientation.
Thirdly, the swimmers can opt to do nothing, that is, only move passively by Brownian motion.

In our investigations, interactions of the particles with their surrounding fluid is modelled by the over-damped Langevin equations
\begin{equation}
      \dot{\vb{r}}_i = \gamma_t^{-1} \left[F(t) \vb{e}_i(\Theta_i) - \nabla V(\vb{r}_i, \{\vb{r}_j\}) \right] + \sqrt{2 k_B T \gamma_t^{-1}} \vb{R}^t_i(t), 
      \label{eq:brownian_pos}
\end{equation}
\begin{equation}
      \dot{\Theta} = \gamma_r^{-1} m(t) + \sqrt{2 k_B T \gamma_r^{-1}} R^r_i(t).
      \label{eq:brownian_angle}
\end{equation}
where, $\vb{r}_i$ is the (two-dimensional) position of particle $i$, $\Theta_i$ the angle describing the particle orientation,  $\gamma_{(t,r)}$ the translational (rotational) friction coefficient, $F$ and $m$ a force and torque corresponding to the respective type of active motion, $\vb{e} = (\cos(\Theta), \sin(\Theta))^T$ the particle orientation, $V$ an interaction potential between all particles in the system, $k_B$ the Boltzmann constant, $T$ the temperature and $\vb{R}^{(t, r)}_i$ a noise term with zero mean and correlations according to $\expval{R^{(t,r)}_i(t) R^{(t,r)}_j(t')} = \delta_{ij} \delta(t-t')$, where $\expval{\cdot}$ denotes an ensemble average.

To quantify the relative importance of active and passive motion, we define translational and rotational P\'{e}clet numbers
\begin{equation}
    Pe^{trans, rot} = \frac{\tscale^{diff}_{trans, rot}}{\tscale^{act}_{trans, rot}}.
\end{equation}
Here,
\begin{equation}
    \tscale_\text{rot}^\text{diff} = \frac{1}{2D_\text{rot}} = \frac{\gamma_r}{2 k_B T}, \quad \tscale_\text{rot}^\text{act} = \frac{2\pi}{\omega^\text{act}},
    \label{eqn:rot-scale}
\end{equation}
are the timescale of decorrelation of the particle director through rotational diffusion and the timescale for one active rotation, respectively.
For the translational degrees of freedom we have
\begin{equation}
    \tscale_\text{trans}^\text{diff} = \frac{a^2}{D_\text{trans}} = \frac{a^2 \gamma_t}{k_B T}, \quad \tscale_\text{trans}^\text{act} = \frac{a}{v^\text{act}},
    \label{eqn:tran-scale}
\end{equation}
as the timescale for diffusion of one particle radius and the timescale for swimming of one particle radius, respectively.
In regimes where $Pe^{trans, rot} \gg 1$ the dynamics will be dominated by active motion and when $Pe^{trans, rot} \ll 1$, it will resemble passive diffusion.

\subsection{Actor-Critic Reinforcement Learning}
Reinforcement learning concerns itself with the interactions between an agent and its environment within which it gradually learns to achieve a desired task~\citep{sutton98a}.
This agent is typically provided with a set of actions it may perform and uses a policy, $\pi(a_{t}|s_{t})$ to decide at time $t$, based on its current state, $s_{t}$, what the best actions, $a_{t}$ will be such that it maximises a reward, $r(s_{t})$.
Over the course of one or many simulations, this policy will be updated so that the agent becomes more efficient at accomplishing this task and maximising its reward,
\begin{equation}
    \centering
    \pi' = \argmax_{\pi} \langle r(s_{t}|\pi) \rangle
\end{equation}
Deep reinforcement learning accomplishes this task using deep neural networks as the policy, $\pi$~\citep{arulkumaran17a}.
During our investigations, the actor-critic approach to deep reinforcement learning has been adopted due to its flexibility and efficacy~\citep{barto83a, grondman12a}.
In actor-critic reinforcement learning, the actor takes on the role of the policy, $\pi_{\theta}$, parameterized by $\theta$, taking as input the current state of the agent and returning oftentimes a distribution over possible actions from which one is selected.
The critic then takes on the role of a value function, $V^{\pi_{\theta}}_{\omega}$, the objective of which is to describe the expected return of an agent starting in state $s_{t}$ and following policy $\pi$.
During training, the actor is tasked with maximising its finite-horizon return of its policy
\begin{equation}
    \centering
    J(\pi_{\theta}) = 
    \left\langle 
        \left [ 
            \sum\limits_{t=0}^{T} \log \pi_{\theta}(a_{t}|s_{t})\cdot A^{\pi_{\theta}}(s_{t}, a_{t})
        \right ]
    \right\rangle_{\tau},
    \label{eqn:update-value}
\end{equation}
where $A^{\pi_{\theta}}$ is the so-called advantage, computed by
\begin{equation}
    \centering
    A^{\pi_{\theta}}_{t} = G(s_{t}, a_{t}) - V^{\pi_{\theta}}_{\omega}(s_{t}),
\end{equation}
where $G$ is an analytic expected returns function.
In our studies, a simple decaying return function
\begin{equation}
    \centering
    G_{t} = \sum\limits_{t'=t}^{T}\epsilon^{t' - t}r_{t'},
\end{equation}
with decay factor $\epsilon$ is used.
$J(\pi_{\theta})$ is maximised by way of gradient ascent on the actor parameters with updates taking the form
\begin{equation}
    \centering
    \theta' = \theta + \eta\cdot\nabla_{\theta}J(\pi_{\theta}),
\end{equation}
with learning rate $\eta$.
Recalling that the actor output is a distribution over actions, should the advantage be negative, i.e, the critic believes a better trajectory could have been chosen, the log probability of these action will be discouraged.
If this number is positive, the actor has outperformed the expectation of the critic and the policy is reinforced.
The critic network is trained directly on the chosen expected returns function via the Huber loss
\begin{equation}
    \centering
    L^{\delta}(y^{\text{true}}, y^{\text{predicted}}) = 
    \begin{cases}
        \frac{1}{2}\cdot(y^{\text{true}} - y^{\text{predicted}})^{2} & \text{, for } | y^{\text{true}} - y^{\text{predicted}}| \leq \delta, \\
        \delta\cdot (|y^{\text{true}} - y^{\text{predicted}}| - \frac{1}{2}\delta) & \text{, otherwise},
    \end{cases}
\end{equation}
with $\delta = 1$ in all studies.
Such an update procedure is referred to as simple or vanilla policy gradient~\citep{sutton99a}.
Whilst more sophisticated approaches exist, namely proximal policy optimization~\citep{schulman17a}, for this particular study, the simpler approach sufficed.

\subsection{Multi-Agent Reinforcement Learning}
In our simulations, we work with not one, but many agents simultaneously, moving from the general concept of reinforcement learning into multi-agent reinforcement learning (MARL)~\citep{gronauer22a}.
In these cases, each agent shares a single actor and critic network and at the update time, also the experience that they have gathered.
During the simulations, each agent asks the actor for an action to take individually and collects its own reward.
At the time of the update, $J(\pi_{\theta})$ becomes,
\begin{equation}
    J_{\text{MARL}} = \frac{1}{N}\sum\limits_{i}^{N} J_{i}(\pi_{\theta}),
\end{equation}
where $i$ sums over the agents in the system and $J_{i}$ is simply Equation~\eqref{eqn:update-value} for a single agent.
In this way, the experience of each agent is accumulated and updated together.

The field of MARL has a a vast set of definitions with respect to how individual agents interact and share knowledge in order to achieve the problem they are training on~\citep{oliehoek16a}.
In this work, a decentralized Markov decision process is used to describe how the agents in the system interact.
The system is considered decentralized as each agent receives only local information regarding its environment and a local reward for its own actions. 
During training, these rewards and local states are summed over and in doing so, the agents share the knowledge with one another.

\section{Methods}
\label{sec:methods}

\subsection{SwarmRL}
\label{subsec:swarmrl}
All reinforcement learning and simulation has been handled through the open-source software package, SwarmRL~\citep{swarmrl}.
SwarmRL is a Python library built to combine molecular dynamics engines with classical control and reinforcement learning algorithms.
All machine learning uses the JAX~\citep{bradbury18a} and Flax~\citep{heek23a} libraries.

\subsection{ESPResSo Simulations}
In this study, reinforcement learning is applied to training microswimmers in a physically realistic simulated environment.
For this environment, we employ the ESPResSo simulation engine~\citep{weik19a}.
Trajectories of the particles are simulated using the over-damped Langevin equations of motion for both position and orientation described in Equations~\eqref{eq:brownian_pos} and ~\eqref{eq:brownian_angle}.
The friction coefficient of a spherical agent with radius $r$ in a fluid with dynamic viscosity $\mu$ is calculated according to Stokes' law as $\gamma_t = 6 \pi \mu r$ and $\gamma_r = 8 \pi \mu r^3$.
Interactions between the spherical agents are modelled with the two-body Weeks-Chandler-Anderson (WCA) potential~\citep{weeks71a}, which can be seen as an almost-hard-sphere interaction
\begin{equation}
    V(r_{ij}) = \begin{cases}
    4\cdot V_{0}\left[ \left(\frac{\sigma}{r_{ij}}\right)^{12} - \left(\frac{\sigma}{r_{ij}}\right)^{6} \right] + V_{0}, \quad &r_{ij}<2^{1/6}\sigma\, \\
    0, &\text{else}.
    \end{cases}
    \label{eqn:wca}
\end{equation}
Here, $r_{ij} = || \vb{r}_i - \vb{r}_j ||_2$ is the Euclidean distance between the particles, and $\sigma = 2a$ the colloid diameter. We choose the interaction strength $V_0 = k_B T$.
Details on the anisotropic particles' friction coefficients and interaction potentials can be found in the supplementary information.
\subsection{Reinforcement Learning Parameters}
\begin{figure}
    \centering
    \includegraphics[width=0.5\linewidth]{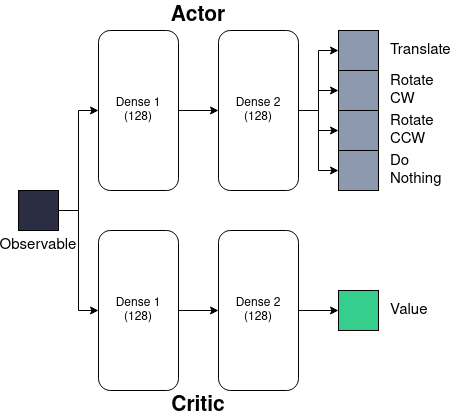}
    \caption{Representation of actor-critic reinforcement learning architectures.}
    \label{fig:ac-arch}
\end{figure}

In our investigations, the actor-critic approach to reinforcement learning is utilised with a network architecture displayed in Figure~\ref{fig:ac-arch}.
A two-layer network, each with 128 units, is deployed for both the actor and the critic, along with ReLU activation functions.
During the training, each network is trained for 10000 episodes, each of which consists of 20 applications of the policy over 4000 simulation time steps. 
Each episode would be \SI{2}{\second} in real-time.
Updates of the network are handled by the Adam optimizer~\citep{kingma17a} using a learning rate of 0.002.
For each swim speed and agent size, 20 reinforcement learning runs were performed to collect statistics.

\subsection{Agent Definition}
\begin{figure}
    \centering
    \includegraphics[width=\linewidth]{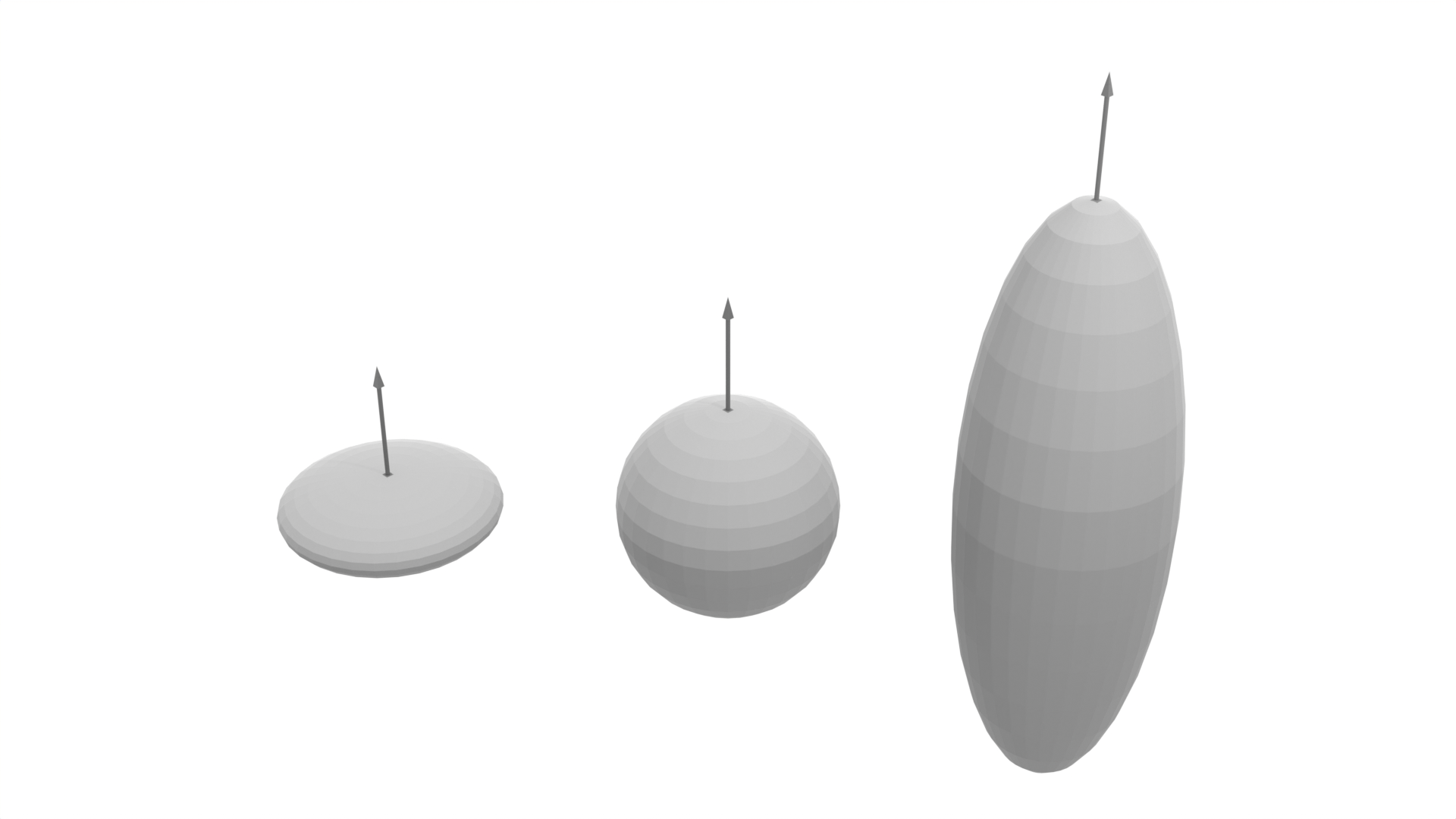}
    \caption{Graphical Representation of the three agent shapes considered in this study, the sphere (center), prolate (right), and oblate (left). In each case, the volume of the agent is kept equal for a given radius value.}
    \label{fig:agents}
\end{figure}

The study considered three agent shapes with a fixed set of actions.
Agent shapes were designed to mimic some of those found in biology: oblate, prolate, and spherical bacteria or microswimmers.
Figure~\ref{fig:agents} displays renderings of these agents for radius $1 \mu m$ constructed using the Vedo Python package~\citep{musy23a}.
As the agents are designed to mimic bacterial particles, we endow them with the ability to perform the four actions described in Section~\ref{sec:bio-chemo},
\begin{equation}
\mathcal{A}=
    \begin{cases}
        \text{Translate:} & v = n\cdot d \mu m s^{-1}\text{, } \omega = 0.0 s^{-1} \\
        \text{Rotate CCW:} & v = 0 \mu m s^{-1} \text{, } \omega =  10.472 s^{-1} \\
        \text{Rotate CW:} & v = 0 \mu m s^{-1} \text{, } \omega = -10.472 s^{-1} \\
        \text{Do Nothing:} & v = 0 \mu m s^{-1} \text{, } \omega = 0.0 s^{-1},
    \end{cases}
\end{equation}
where $d$ is the colloid diameter, $n$ is a scaling factor that we vary during the experiment, and $\omega$ is the angular velocity measured in radians per second.
The rotation speed was chosen to be similar to that of Escherichia coli~\citep{berg72a}.
In line with~\citep{murray92a}, we argue that agent volume is proportional to its swimming speed. 
Therefore, the action is measured in body lengths, and all agents with the same radius will swim at the same speed.
The agents receive a state description designed to resemble a bacterium sensing changes in its surroundings, defined mathematically by
\begin{equation}
    \centering
    o_{i}(t) = f\left(||\mathbf{r}_{i}(t) - \mathbf{r}_{s}(t)||_{2}\right) - f\left(||\mathbf{r}_{i}(t-\Delta t) - \mathbf{r}_{s}(t-\Delta t)||_{2}\right),
    \label{eqn:concentration-sensing}
\end{equation}
where $o_{i}$ is the observable for the $i^\text{th}$ agent, $f$ is the field chosen to represent the chemical being sensed, in our study, $\frac{1}{r_{is}}$,  $\mathbf{r}_{i}(t)$ is the position of the $i^{\text{th}}$ agent at time $t$, $\Delta t$ is the amount of time since the last action was computed and $\hat{\mathbf{r}}_{s}(t)$ denotes the position of the source of the field at time $t$.
To encourage chemotaxis, agents are rewarded using a similar function
\begin{equation}
    \centering
    r_{i}(t) = 
    \begin{cases}
        o_{i} & \text{ if } o_{i} > 0 \\
        0 & \text{ else.}
    \end{cases}
\end{equation}
This way, movement towards the source is encouraged, but movement away is not explicitly discouraged. 
We further refrain from using an absolute measure of the field in this study as it would not resemble the natural sensing abilities of the bacteria~\citep{bren00a}.
The addition of such a reward might be used to encourage agents to form groups reminiscent of biofilms or to replicate the act of digesting the source of the field, both of which are left for future studies.

\subsection{Computational Methods}
Training and deployment of the reinforcement learning models was performed on the University of Stuttgart SimTech compute cluster.
Each simulation and training routine utilised six threads of an AMD EPYC 7702 CPU node, and all simulations were run in parallel.
Due to the system sizes and machine learning being performed, no GPUs were required for these experiments.
Training of each model required approximately twenty-four hours, and the deployment simulations were approximately six hours.
The simulations and models were analysed on the same cluster hardware.

\section{Results}
\label{sec:results}
Due to the similarity in the results and the amount of analysis, only the plots for the spherical agent analysis are shown in the main manuscript.
All other plots are included in the SI, and any deviations between results are mentioned here.
\subsection{Probability of Emergent Chemotaxis}
This investigation aims to identify limits on emergent chemotaxis in RL agents in the hope that such limits cross over into biology, allowing us to study natural biological processes using RL as a valid surrogate model.
These limits suggest formulating a phase diagram with forbidden regions where this behaviour is strictly prohibited. 
To this end, all simulations were collected where the final 50 \% of the deployment trajectory was below \SI{15}{\micro\meter} from the source of the field.
This distance was determined based on the visual observation that no model that had successfully learned chemotaxis was farther away from the source than this distance.
The successful simulations were used to compute the probability of learning chemotaxis by rationing them against the total number of simulations performed for a single speed and agent size.
\begin{figure}[htbp]
    \centering
    \includegraphics[width=\linewidth]{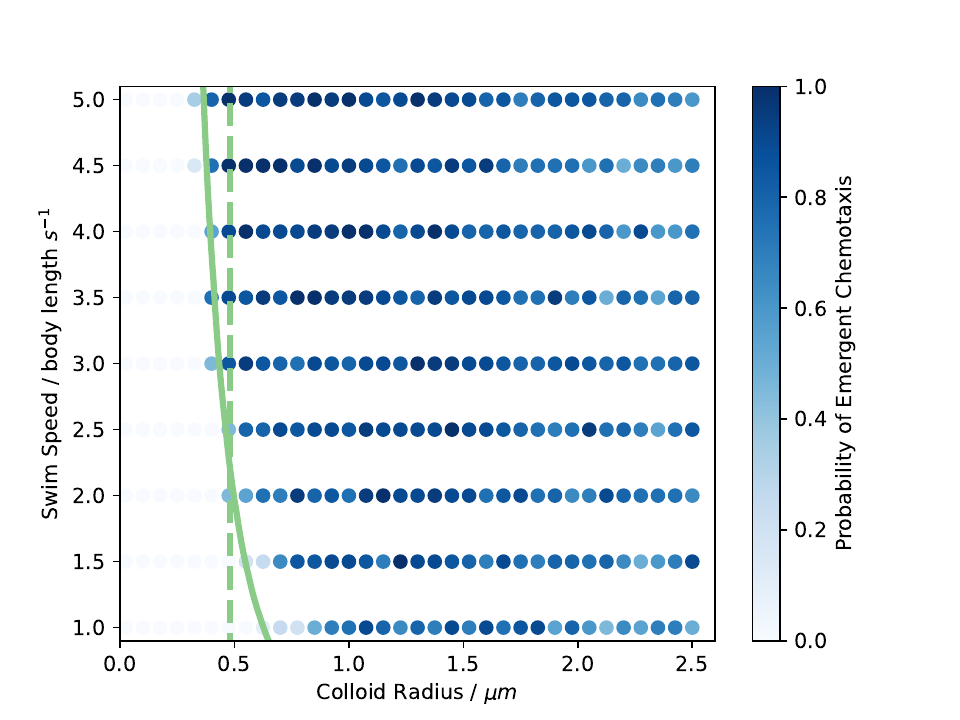}
    \caption{Probability of successful chemotaxis emerging from RL studies. Raw data from the experiment. The colour of each point corresponds to the number of RL simulations that successfully learned how to perform chemotaxis. The green lines indicate the theoretical values at which translational (solid) and rotational (dashed) diffusion becomes dominant compared to the active motion of the agents.}
    \label{fig:phase-diagram}
\end{figure}
Figure~\ref{fig:phase-diagram} shows the computed phase diagram.
The figure plots the sampled data points; the alpha value corresponds to the probability, with the more transparent points being less probable.
One can identify a forbidden region in the size-speed space.
Interestingly, it appears that smaller, faster colloids are more likely to learn an effective chemotaxis policy.
Suppose we consider the difficulty the RL algorithm has in training a policy with the real-world problem of evolving a suitable structure for life. 
In that case, these results suggest a trade-off between speed and size when learning how to perform chemotaxis.
The most critical component of Figure~\ref{fig:phase-diagram} is the theoretical boundaries formed by considering the ratio between Brownian motion and the active motion of the particles described by Equations~\ref{eqn:rot-scale} and~\ref{eqn:tran-scale}.
The green lines in Figure~\ref{fig:phase-diagram} correspond to the colloid radius and speed values for which this ratio is $1.0$ for translational (solid) and rotational (dashed) diffusion.
The translation ratio forms a boundary where the RL agents can no longer learn successful chemotaxis.
The rotational diffusion line appears less strict, particularly for the faster agents, where it appears that with enough translational activity, the agents can overcome having rotational diffusion dominate over active rotation.
The alignment of our results with the theoretical values suggests that it does so as soon as it is physically possible for an RL agent to learn chemotaxis.
Such a result encourages one to consider studying further features of the models to understand how these features might also arise in these agents' biological or artificial counterparts.
Interestingly, the onset of successful chemotaxis can take place far below this theoretical limit but very rarely.

\subsection{Learning Efficiency}
\begin{figure}[htbp]
    \centering
    \includegraphics[width=\linewidth]{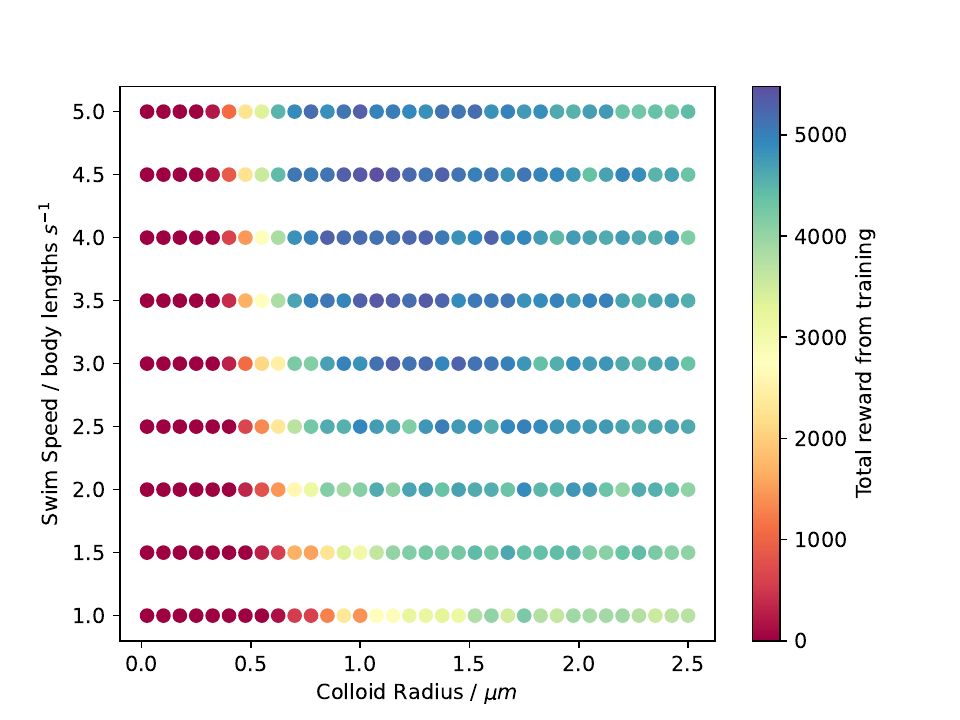}
    \caption{Probability of successful chemotaxis emerging from RL studies. Raw data from the experiment. The colour of each point corresponds to the maximum reward achieved by the agents during the 10'000 episodes. }
    \label{fig:learning-diagram}
\end{figure}

Next, we look at how the reward received from the reinforcement learning process changed depending on the size and speed of the colloids.
This measure will indicate how easy it was for the model to learn the policy required to perform chemotaxis.
Figure~\ref{fig:learning-diagram} outlines the results of this study in a similar manner to Figure~\ref{fig:phase-diagram}.
In the figure, the point's colour corresponds to the total reward accumulated by the agents during all 10'000 training episodes.
In order to compute the colour values in Figure~\ref{fig:learning-diagram}, we corrected the size difference between the colloids.
In the original simulations, an explicit distance to the source is used in the reward computation. 
However, this biases the results such that smaller colloids, no matter how successful they were, will achieve exponentially higher rewards as they can approach the source more closely. 
Therefore, the rewards in Figure~\ref{fig:learning-diagram} were computed by converting the reward from distance to the number of body lengths from the source.
We can see that the reward diagrams roughly mirror the results shown in Figure~\ref{fig:phase-diagram} with larger discrepancies between the larger and smaller colloids.
Namely, the rewards achieved by small and fast agents is noticeably larger than those of the bigger agents.
This effect is particularly evident in the prolate and oblate simulations (SI Figures~\ref{fig:oblate-reward} and~\ref{fig:prolate-rew}). 
It is likely due to their hopping over the intended target and inability to sit on it as accurately as the smaller agents.

\subsection{Policy Efficiency}
\begin{figure}[htbp]
     \centering
     \includegraphics[width=\linewidth]{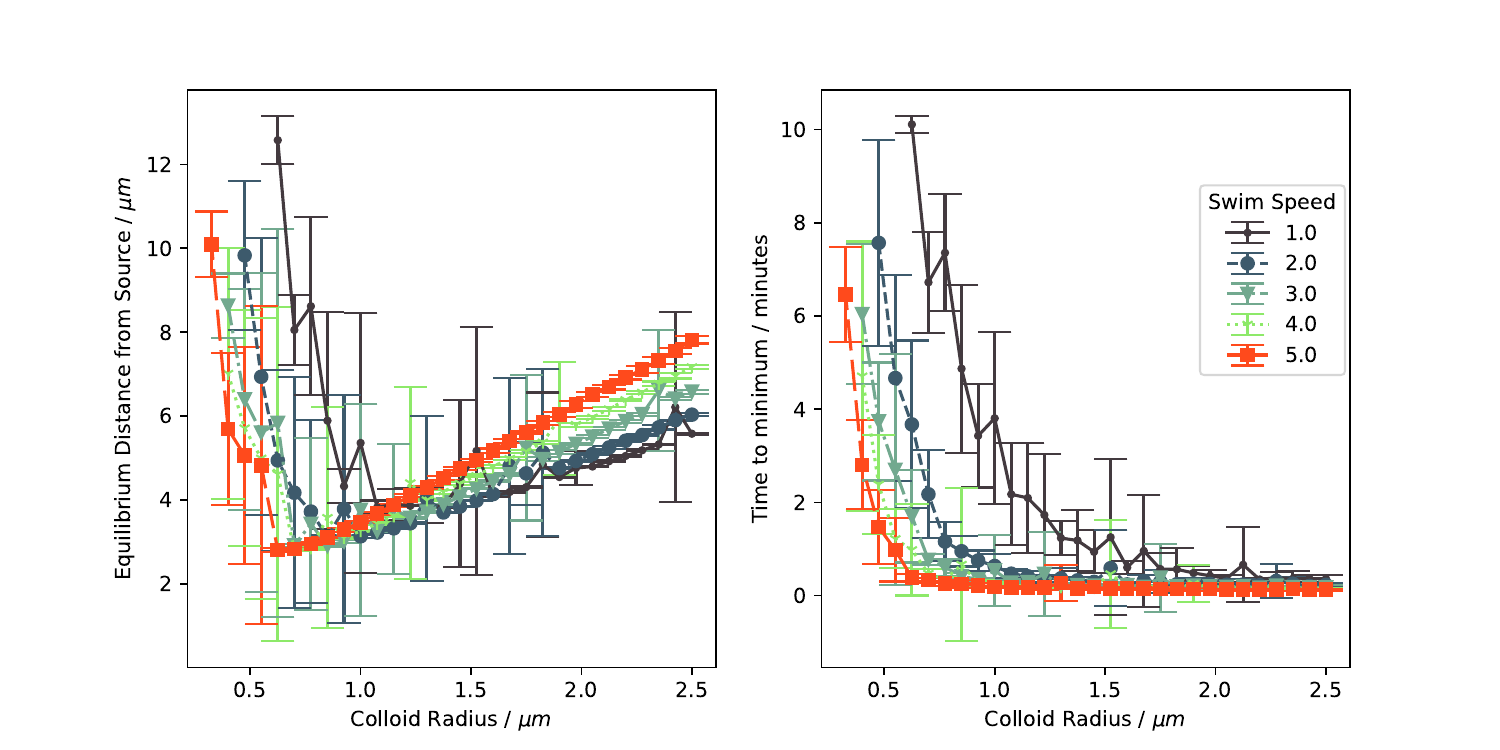}
        \caption{(left) Mean distance from the source for each swim speed and colloid size. A clear minimum in each plot suggests an optimal size dependent on swim speed. (right) Rate of convergence to the source for different swim speeds and sizes. Interestingly, the convergence rate of larger colloids is relatively similar, suggesting some redundancy in larger body sizes and swim speeds.}
        \label{fig:policy-efficiency}
\end{figure}

It is clear from the previous section that microswimmers of different sizes and speeds differ in their probability of emergent chemotaxis. 
However, what the differences are, if any, between their adopted strategy still needs to be determined.
To identify these differences, the deployment simulations were analysed to compute the final equilibrium distance of the colloids around the source as well as after how many action updates they reached this distance.
Figure~\ref{fig:policy-efficiency} displays the results of this investigation.
Figure~\ref{fig:policy-efficiency} (left) details the equilibrium distance of the agents as a function of radius for all studied swim speeds.
On the right-hand side, we see a clear emergence of a linear trend as the limiting factor getting closer to the source becomes the size of the colloid and its speed. 
The left non-linear side of the plot also contains interesting features. 
Aside from the speed of one, the minimum distance from the source of the chemical field is achieved at a similar colloid size for all of the different speeds, with faster agents able to achieve slightly better equilibrium distances with smaller bodies. 
In Figure~\ref{fig:policy-efficiency} (right), we see the average time the colloids reach the equilibrium distance.
For the smaller agents, as is perhaps intuitive, the faster colloids can orient and move themselves to the source faster than their slower counterparts. 
However, this relationship fades for larger colloids as we see that after approximately \SI{1}{\micro\meter} radii, all colloid sizes and speeds converge at similar times except for the slowest at one body length. 
The time to minimum converges slightly above \SI{25}{\second}.
The results also suggest that after \SI{0.5}{\micro\meter} radius, there is no conceivable benefit and, in fact, due to the larger equilibrium distance, perhaps even a detriment in being larger. 
Interestingly, the most unstable equilibrium distances, identified by large variance in mean value and distance from the source, occur close to or within the region displayed in Figure~\ref{fig:phase-diagram} where rotational diffusion overpowers the active rotation of the agents.
This strong environmental effect could cause instability in these models as they must rely solely on their active translation to achieve chemotaxis.

\subsection{Emergent Policy}
\begin{figure}[htbp]
    \centering
    \includegraphics[width=0.8\linewidth]{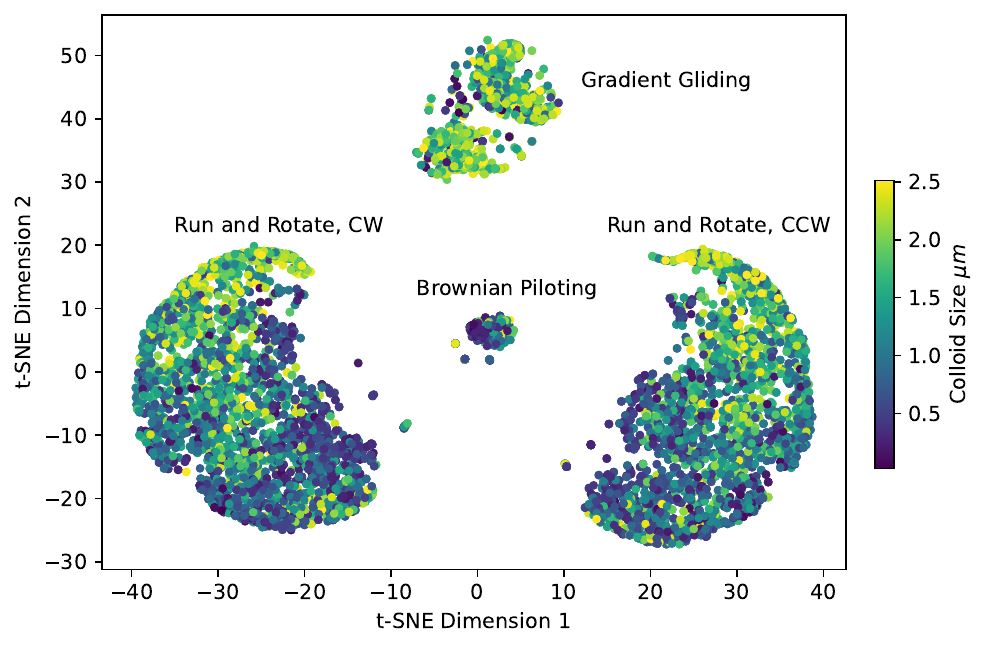}
    \caption{t-SNE embedding of the policy vectors for all successful agents in the study. Four large groups are formed, corresponding to the policies learned by the agents. The colour in these diagrams corresponds to the size of the studied agents.}
    \label{fig:tsne}
\end{figure}

As a final investigation, we determine whether the emergent policy of the RL agent differs for changing physical properties and shapes.
Studying the particles' trajectory alone is almost impossible as they do not show large deviations from one another.
Therefore, to do so, the trained models are given test data over a domain, $x \in (-10.0, 10.0)$ and the probability of selecting each action is computed from the network outputs.
The network output will be four numbers for each concentration value; therefore, before performing further analysis, these outputs are flattened into a single vector for each model, which we will refer to as the probability vectors of the network.
In order to identify any structure in the data, we study the two-component t-distributed stochastic neighbourhood embedding (t-SNE)~\citep{maaten08a} of the probability vectors as implemented in the sklearn Python package~\citep{pedregosa11a}.
Figure~\ref{fig:tsne} outlines the results of the t-SNE for the policy data with a perplexity of 300 and principle-component-analysis (PCA) initialisation.
Examining the t-SNE plots, we see the emergence of four groups, one of which is seemingly divided into two smaller subgroups.
Using this information, we perform k-means clustering~\citep{lloyd82a} on the probability vectors to split them into four clusters.
The probability of the outcome of each policy is listed in Table~\ref{tab:statistics} along with the explained variance from a PCA decomposition of the probability vectors.
The probabilities are computed by examining the number of points clustered into each class by the k-means algorithm, which we assign to a policy by directly examining the action probabilities of the agents mapping into the class.
The diagrams used to perform this mapping are included in the SI, where we show the probabilities of each action being taken for all agent sizes, shapes, and speeds.
We also include smaller sample policy diagrams to demonstrate the actions taken by the agents for each strategy in SI Figure 7.
However, as we only ask for four classes, this percentage will naturally ignore more convoluted and rare policies that the t-SNE and K-Means cannot sufficiently distinguish from others.
We also perform PCA decomposition on the probability vectors to identify how much each policy explains the data distribution.
In this approach, we see that 5.9 \% of the data belongs to components with a smaller than one \% impact on the variance of the PCA.
When we examine the policies in this region, they are typically made up of weak combinations of the more dominant policies with very few exceptions.
These policies may explain the splitting in the medium-sized group in Figure~\ref{fig:tsne}.
\begin{table}[htbp]
\centering
\begin{tabular}{@{}lcc@{}}
\toprule
Policy Name                       & \multicolumn{1}{l}{Percentage Learned (K-Means)} & \multicolumn{1}{l}{Explained Variance (PCA)} \\ \midrule
Run and Rotate          & 83.49                                            & 83.5                                         \\
Gradient Gliding & 12.88                                            & 7.1                                          \\
Brownian Piloting           & 3.63                                             & 3.5                                          \\
Exotic Policies                      & 0.0                                              & 5.9                                          \\ \bottomrule
\end{tabular}
\caption{Percentage and explained variance of agents which learned specific policy along with the explained variance of the principle components for each policy identified.}
\label{tab:statistics}
\end{table}
The remainder of the section will discuss each emergent policy in detail, including some policies poorly captured by the embedding methods.
\paragraph{Run and Rotate}
In the vast majority of cases (83 \%), the agents learned a policy strikingly similar to the \texttt{run-and-tumble} approach found in nature.
In these cases, upon experiencing a negative input to the network, signifying a movement away from the source of the gradient, the agents rotate either CW or CCW.
Interestingly, once the agents chose a direction to rotate, they did not use the other one.
Upon positive input to the network, i.e., movement towards the source of the gradient, the agents chose to translate with probability 1. 
This policy can be seen clearly for the larger colloids in the top two rows of SI Figure~\ref{fig:small-policy-examples}.
CW vs CCW selection was even throughout the simulations, with no preferred direction discovered.

\paragraph{Gradient Gliding}
For large colloids, some took translate for most inputs, only rotating for minimal changes in the gradient, such as those occurring when far away from the source or moving equipotential around it.
Even in these cases, the strategy is inconsistent and still has a high probability of translation.
While this strategy might appear strange initially, it helps to consider regions where the gradient will be so slight.
As the agents are initialised in the simulation, they will sit in a region with minimal gradient.
Upon moving around this area, they will likely rotate until they are dragged into a region where the gradient increases enough for it to begin translating.
This was common amongst the chosen policies, with somewhere between 7 \% and 12 \% opting for this approach.
The discrepancy in percentage arises due to the spurious policies discussed in a later section.
We label this policy \texttt{Gradient Gliding} as the colloids generally follow a translation path with very small adjustments made under low gradient changes.
An example plot of this strategy can be seen in SI Figure~\ref{fig:small-policy-examples}, row three.

\paragraph{Brownian Piloting}
An alternative policy, referred to here as \texttt{Brownian Piloting}, was seen particularly in the case of smaller agents where rotational Brownian motion overcomes the active rotation.
The agents learned to do nothing when experiencing negative network inputs and to translate if they see a positive one.
In this way, the agents do not fight against the Brownian forces when their active motion cannot overcome it.
As this was only seen in the small agents, it is clear that such a policy could be more optimal. 
However, it does demonstrate that small, weak agents can still successfully learn to navigate toward sources of nutrition.
Overall, this policy was adopted in 3.6 \% of cases.
One can see Brownian Piloting in the fourth row of SI Figure~\ref{fig:small-policy-examples}.

\paragraph{Exotic policies (EP)}
As was previously mentioned, approximately 5.9 \% of the emergent policies were not well mapped into single classes.
However, we can identify several so-called exotic policies by manually looking at the action probabilities.
\begin{itemize}
    \item EP 1: In these cases, the agents only translated when their input was negative. 
Otherwise, they chose to do nothing. 
        This was only observed in small agents where random fluctuations significantly impacted their rotation more than active swimming.
    \item EP 2: In this approach, we saw the agents choose the \texttt{Do Nothing} action at almost all times, except when the input to the network was a small positive gradient, at which point they would translate.
    This approach occurred for small colloids where all random forces outnumbered their active swimming. 
    This indicates that microswimmers can survive in cases where their swimming is overpowered, effectively using their environment to perform successful chemotaxis.
    Whilst extremely inefficient, this swim strategy preserves energy and successfully allows the agents to perform chemotaxis despite an almost impossible condition.
    \item Combinations: We noticed combinations of other, more common approaches in many exotic policies.
    For example, some smaller colloids in Brownian-dominated regimes performed active rotation in both CW and CCW directions for negative inputs and translate for positive. 
    We identify this policy as more or less equivalent to \texttt{Brownian Piloting} as the active rotation will not yield more than simply sitting still.
    In other cases, the onset of translation was delayed or accelerated, yielding slight variations of \texttt{Run and Rotate}.
    As these points combined mixtures of the more dominant policies and yet did not occur often, the clustering algorithms could not successfully separate them into distinct classes.
\end{itemize}
The results tell us that in the cases where active translation and rotation are possible and dominate over Brownian effects, the agents often learn to perform a run-and-rotate trajectory.
In cases where Brownian effects dominate active motion, the agents learn to adapt to this environment by performing only actions that move them into a new environment, by using the Brownian forces to their advantage.
Interestingly, these policies are not well differentiated within the trajectories alone; only by looking at the neural networks can we see how the agents make decisions.
Such an insight might guide us in understanding how real-world bacteria navigate their environments, and perhaps, how to disrupt, as in quorum quenching~\citep{grandclement15a}, or support, as in quorum enhancement~\citep{garcia14a}, this navigation.

\section{Conclusion}
\label{sec:conclusion}
In this study, we have tested the role of size and swim speed on the emergent strategy of microscopic active agents learning chemotaxis via multi-agent actor-critic reinforcement learning.
Our simulations demonstrated that intelligent agents can learn chemotactic behaviour, even in environments where Brownian random forces begin to dominate their active motion.
In such regimes, we found that the chemotaxis was not optimal in terms of their equilibrium distance from the source of the chemical gradient or the speed at which they made it to the source. 
However, they could consistently reach their target.
Interestingly, we saw that as the P\'{e}clet number grew above one and active motion dominated the Brownian forces and torques, the learned policies also converged quickly to a similar equilibrium distance and time.
After studying the policy efficiency, we looked into the strategies adopted by the agents to perform chemotaxis.
We identified three dominant strategies which we named \texttt{Run and Rotate}, \texttt{Gradient Gliding}, and \texttt{Brownian Piloting}.
The first policy occurred most often (83.5 \%) but predominantly in cases where the colloids were large enough to no longer be in regions where rotational or translation Brownian motion overcame their active swimming.
This strategy involved translating as long as the input to the agent was positive, i.e., moving towards the source and rotating if it was negative.
The second most common strategy (~7.1 \% emergence), also occurring in larger colloids, was to translate for most of the time whether the input was negative or positive, and only when the input to the network was small would they sometimes choose to rotate.
This strategy meant that the colloids spent a long time rotating when they were far away from the source but translating when they identified the direction of the source.
The final common strategy occurred when the rotational and sometimes translational Brownian motion dominated the active swimming.
In these cases, the agents would perform the \texttt{Do Nothing} action while the input to the network was negative, and only when it was positive would they begin translating.
We identified further policies, including a kind of \texttt{lazy swimming} where the agents performed no actions except when they were far away from the source and the input to the network was weakly positive, as well as some cases where agents learned to rotate in both CW and CCW directions but often in a regime where this was not useful.
Overall, we have identified that reinforcement learning can replicate natural behaviour of organisms. 
However, it can also provide insight into biological swimmers' possible strategies and may provide a path forward for exploiting this knowledge.
A further point of interest would be to identify natural biological swimmers who have evolved such swimming patterns or can outperform the emergent strategies of the RL agents.

\section{Data and Availability}
All data can be made available upon reasonable request to the authors and, upon publication, will be made publicly available through the DaRUS service.

\section{Acknowledgements}
C.H and S.T acknowledge financial support from the German Funding Agency (Deutsche Forschungsgemeinschaft DFG) under Germany’s Excellence Strategy EXC 2075-390740016, and S. T was supported by an LGF stipend of the state of Baden-W\"{u}rttemberg.
C.H and S.T acknowledge financial support from the German Funding Agency (Deutsche Forschungsgemeinschaft DFG) under the Priority Program SPP 2363.
C.H and C.L acknowledge funding by the Deutsche Forschungsgemeinschaft (DFG, German Research Foundation) under Project Number 327154368-SFB 1313.
The authors would like to acknowledge funding from the Deutsche Forschungsgemeinschaft (DFG, German Research Foundation) Compute Cluster grant no. 492175459.

\bibliography{bibliography}

\begin{thebibliography}{40}
\providecommand{\natexlab}[1]{#1}
\providecommand{\url}[1]{\texttt{#1}}
\expandafter\ifx\csname urlstyle\endcsname\relax
  \providecommand{\doi}[1]{doi: #1}\else
  \providecommand{\doi}{doi: \begingroup \urlstyle{rm}\Url}\fi

\bibitem[Arulkumaran et~al.(2017)Arulkumaran, Deisenroth, Brundage, and
  Bharath]{arulkumaran17a}
K.~Arulkumaran, M.~P. Deisenroth, M.~Brundage, and A.~A. Bharath.
\newblock Deep reinforcement learning: A brief survey.
\newblock \emph{IEEE Signal Processing Magazine}, 34\penalty0 (6):\penalty0
  26--38, 2017.
\newblock \doi{10.1109/MSP.2017.2743240}.

\bibitem[Barto et~al.(1983)Barto, Sutton, and Anderson]{barto83a}
A.~G. Barto, R.~S. Sutton, and C.~W. Anderson.
\newblock Neuronlike adaptive elements that can solve difficult learning
  control problems.
\newblock \emph{IEEE Transactions on Systems, Man, and Cybernetics},
  SMC-13\penalty0 (5):\penalty0 834--846, 1983.
\newblock \doi{10.1109/TSMC.1983.6313077}.

\bibitem[Bastos-Arrieta et~al.(2018)Bastos-Arrieta, Revilla-Guarinos, Uspal,
  and Simmchen]{bastos18a}
J.~Bastos-Arrieta, A.~Revilla-Guarinos, W.~E. Uspal, and J.~Simmchen.
\newblock Bacterial biohybrid microswimmers.
\newblock \emph{Frontiers in Robotics and AI}, 5, 2018.
\newblock ISSN 2296-9144.
\newblock \doi{10.3389/frobt.2018.00097}.
\newblock URL
  \url{https://www.frontiersin.org/articles/10.3389/frobt.2018.00097}.

\bibitem[Berg(2004)]{berg04a}
H.~Berg.
\newblock coli in motion2004springer.
\newblock \emph{New York}, 2004.

\bibitem[Berg and Brown(1972)]{berg72a}
H.~C. Berg and D.~A. Brown.
\newblock Chemotaxis in escherichia coli analysed by three-dimensional
  tracking.
\newblock \emph{Nature}, 239\penalty0 (5374):\penalty0 500--504, Oct 1972.
\newblock ISSN 1476-4687.
\newblock \doi{10.1038/239500a0}.
\newblock URL \url{https://doi.org/10.1038/239500a0}.

\bibitem[Bradbury et~al.(2018)Bradbury, Frostig, Hawkins, Johnson, Leary,
  Maclaurin, Necula, Paszke, Vander{P}las, Wanderman-{M}ilne, and
  Zhang]{bradbury18a}
J.~Bradbury, R.~Frostig, P.~Hawkins, M.~J. Johnson, C.~Leary, D.~Maclaurin,
  G.~Necula, A.~Paszke, J.~Vander{P}las, S.~Wanderman-{M}ilne, and Q.~Zhang.
\newblock {JAX}: composable transformations of {P}ython+{N}um{P}y programs,
  2018.
\newblock URL \url{http://github.com/google/jax}.

\bibitem[Bren and Eisenbach(2000)]{bren00a}
A.~Bren and M.~Eisenbach.
\newblock How signals are heard during bacterial chemotaxis: protein-protein
  interactions in sensory signal propagation.
\newblock \emph{J Bacteriol}, 182\penalty0 (24):\penalty0 6865--6873, Dec.
  2000.

\bibitem[Darnton et~al.(2006)Darnton, Turner, Rojevsky, and Berg]{darnton06a}
N.~C. Darnton, L.~Turner, S.~Rojevsky, and H.~C. Berg.
\newblock On torque and tumbling in swimming escherichia coli.
\newblock \emph{J Bacteriol}, 189\penalty0 (5):\penalty0 1756--1764, Dec. 2006.

\bibitem[Datta and Srivastava(1999)]{datta99a}
S.~Datta and D.~K. Srivastava.
\newblock Stokes drag on axially symmetric bodies: a new approach.
\newblock \emph{Proceedings - Mathematical Sciences}, 109\penalty0
  (4):\penalty0 441--452, Nov 1999.
\newblock ISSN 0973-7685.
\newblock \doi{10.1007/BF02838005}.
\newblock URL \url{https://doi.org/10.1007/BF02838005}.

\bibitem[Elgeti et~al.(2015)Elgeti, Winkler, and Gompper]{elgeti15a}
J.~Elgeti, R.~G. Winkler, and G.~Gompper.
\newblock Physics of microswimmers—single particle motion and collective
  behavior: a review.
\newblock \emph{Reports on Progress in Physics}, 78\penalty0 (5):\penalty0
  056601, apr 2015.
\newblock \doi{10.1088/0034-4885/78/5/056601}.
\newblock URL \url{https://dx.doi.org/10.1088/0034-4885/78/5/056601}.

\bibitem[Garc{\'\i}a-Contreras et~al.(2014)Garc{\'\i}a-Contreras,
  Nu{\~n}ez-L{\'o}pez, Jasso-Ch{\'a}vez, Kwan, Belmont, Rangel-Vega, Maeda, and
  Wood]{garcia14a}
R.~Garc{\'\i}a-Contreras, L.~Nu{\~n}ez-L{\'o}pez, R.~Jasso-Ch{\'a}vez, B.~W.
  Kwan, J.~A. Belmont, A.~Rangel-Vega, T.~Maeda, and T.~K. Wood.
\newblock Quorum sensing enhancement of the stress response promotes resistance
  to quorum quenching and prevents social cheating.
\newblock \emph{ISME J}, 9\penalty0 (1):\penalty0 115--125, June 2014.

\bibitem[Gay and Berne(1981)]{gay81a}
J.~G. Gay and B.~J. Berne.
\newblock {Modification of the overlap potential to mimic a linear site–site
  potential}.
\newblock \emph{The Journal of Chemical Physics}, 74\penalty0 (6):\penalty0
  3316--3319, 03 1981.
\newblock ISSN 0021-9606.
\newblock \doi{10.1063/1.441483}.
\newblock URL \url{https://doi.org/10.1063/1.441483}.

\bibitem[Grandclément et~al.(2015)Grandclément, Tannières, Moréra, Dessaux,
  and Faure]{grandclement15a}
C.~Grandclément, M.~Tannières, S.~Moréra, Y.~Dessaux, and D.~Faure.
\newblock {Quorum quenching: role in nature and applied developments}.
\newblock \emph{FEMS Microbiology Reviews}, 40\penalty0 (1):\penalty0 86--116,
  10 2015.
\newblock ISSN 0168-6445.
\newblock \doi{10.1093/femsre/fuv038}.
\newblock URL \url{https://doi.org/10.1093/femsre/fuv038}.

\bibitem[Gronauer and Diepold(2022)]{gronauer22a}
S.~Gronauer and K.~Diepold.
\newblock Multi-agent deep reinforcement learning: a survey.
\newblock \emph{Artificial Intelligence Review}, 55\penalty0 (2):\penalty0
  895--943, Feb 2022.
\newblock ISSN 1573-7462.
\newblock \doi{10.1007/s10462-021-09996-w}.
\newblock URL \url{https://doi.org/10.1007/s10462-021-09996-w}.

\bibitem[Grondman et~al.(2012)Grondman, Busoniu, Lopes, and
  Babuska]{grondman12a}
I.~Grondman, L.~Busoniu, G.~A.~D. Lopes, and R.~Babuska.
\newblock A survey of actor-critic reinforcement learning: Standard and natural
  policy gradients.
\newblock \emph{IEEE Transactions on Systems, Man, and Cybernetics, Part C
  (Applications and Reviews)}, 42\penalty0 (6):\penalty0 1291--1307, 2012.
\newblock \doi{10.1109/TSMCC.2012.2218595}.

\bibitem[Hansen et~al.(2007)Hansen, Endres, and Wingreen]{hansen07a}
C.~H. Hansen, R.~G. Endres, and N.~S. Wingreen.
\newblock Chemotaxis in escherichia coli: a molecular model for robust precise
  adaptation.
\newblock \emph{PLoS Comput Biol}, 4\penalty0 (1):\penalty0 e1, Nov. 2007.

\bibitem[Hartl et~al.(2021)Hartl, Hübl, Kahl, and Zöttl]{hartl21a}
B.~Hartl, M.~Hübl, G.~Kahl, and A.~Zöttl.
\newblock Microswimmers learning chemotaxis with genetic algorithms.
\newblock \emph{Proceedings of the National Academy of Sciences}, 118\penalty0
  (19):\penalty0 e2019683118, 2021.
\newblock \doi{10.1073/pnas.2019683118}.
\newblock URL \url{https://www.pnas.org/doi/abs/10.1073/pnas.2019683118}.

\bibitem[Heek et~al.(2023)Heek, Levskaya, Oliver, Ritter, Rondepierre, Steiner,
  and van {Z}ee]{heek23a}
J.~Heek, A.~Levskaya, A.~Oliver, M.~Ritter, B.~Rondepierre, A.~Steiner, and
  M.~van {Z}ee.
\newblock {F}lax: A neural network library and ecosystem for {JAX}, 2023.
\newblock URL \url{http://github.com/google/flax}.

\bibitem[Kingma and Ba(2017)]{kingma17a}
D.~P. Kingma and J.~Ba.
\newblock Adam: A method for stochastic optimization, 2017.

\bibitem[Koenig(1975)]{koenig75a}
S.~H. Koenig.
\newblock Brownian motion of an ellipsoid. a correction to perrin's results.
\newblock \emph{Biopolymers}, 14\penalty0 (11):\penalty0 2421--2423, 1975.
\newblock \doi{https://doi.org/10.1002/bip.1975.360141115}.
\newblock URL
  \url{https://onlinelibrary.wiley.com/doi/abs/10.1002/bip.1975.360141115}.

\bibitem[Lloyd(1982)]{lloyd82a}
S.~Lloyd.
\newblock Least squares quantization in pcm.
\newblock \emph{IEEE Transactions on Information Theory}, 28\penalty0
  (2):\penalty0 129--137, 1982.
\newblock \doi{10.1109/TIT.1982.1056489}.

\bibitem[Mo and Bian(2022)]{mo22a}
C.~Mo and X.~Bian.
\newblock Chemotaxis of sea urchin sperm cells through deep reinforcement
  learning, 2022.
\newblock URL \url{https://arxiv.org/abs/2209.07407}.

\bibitem[Muiños-Landin et~al.(2021)Muiños-Landin, Fischer, Holubec, and
  Cichos]{landin21a}
S.~Muiños-Landin, A.~Fischer, V.~Holubec, and F.~Cichos.
\newblock Reinforcement learning with artificial microswimmers.
\newblock \emph{Science Robotics}, 6\penalty0 (52):\penalty0 eabd9285, 2021.
\newblock \doi{10.1126/scirobotics.abd9285}.
\newblock URL
  \url{https://www.science.org/doi/abs/10.1126/scirobotics.abd9285}.

\bibitem[Murray and Jackson(1992)]{murray92a}
A.~G. Murray and G.~A. Jackson.
\newblock Viral dynamics: a model of the effects of size, shape, motion and
  abundance of single-celled planktonic organisms and other particles.
\newblock \emph{Marine Ecology Progress Series}, 89\penalty0 (2/3):\penalty0
  103--116, 1992.
\newblock ISSN 01718630, 16161599.
\newblock URL \url{http://www.jstor.org/stable/24831780}.

\bibitem[Musy et~al.(2023)Musy, Jacquenot, Dalmasso, Lee, de~Bruin, Soltwedel,
  Tulldahl, Zhou, RobinEnjalbert, Pollack, Hacha, Claudi, Badger, Lu, Sol,
  Yershov, Sullivan, Lerner, Hrisca, Volpatto, Evan, Matzkin, JohnsWor,
  mkerrinrapid, Schlömer, RichardScottOZ, and Schneider]{musy23a}
M.~Musy, G.~Jacquenot, G.~Dalmasso, J.~Lee, R.~de~Bruin, J.~Soltwedel,
  M.~Tulldahl, Z.-Q. Zhou, RobinEnjalbert, A.~Pollack, B.~Hacha, F.~Claudi,
  C.~Badger, X.~Lu, A.~Sol, A.~Yershov, B.~Sullivan, B.~Lerner, D.~Hrisca,
  D.~Volpatto, Evan, F.~Matzkin, JohnsWor, mkerrinrapid, N.~Schlömer,
  RichardScottOZ, and O.~Schneider.
\newblock marcomusy/vedo: 2023.5.0, Nov. 2023.
\newblock URL \url{https://doi.org/10.5281/zenodo.4587871}.

\bibitem[Oliehoek et~al.(2016)Oliehoek, Amato, et~al.]{oliehoek16a}
F.~A. Oliehoek, C.~Amato, et~al.
\newblock \emph{A concise introduction to decentralized POMDPs}, volume~1.
\newblock Springer, 2016.

\bibitem[Pedregosa et~al.(2011)Pedregosa, Varoquaux, Gramfort, Michel, Thirion,
  Grisel, Blondel, Prettenhofer, Weiss, Dubourg, Vanderplas, Passos,
  Cournapeau, Brucher, Perrot, and Duchesnay]{pedregosa11a}
F.~Pedregosa, G.~Varoquaux, A.~Gramfort, V.~Michel, B.~Thirion, O.~Grisel,
  M.~Blondel, P.~Prettenhofer, R.~Weiss, V.~Dubourg, J.~Vanderplas, A.~Passos,
  D.~Cournapeau, M.~Brucher, M.~Perrot, and E.~Duchesnay.
\newblock Scikit-learn: Machine learning in {P}ython.
\newblock \emph{Journal of Machine Learning Research}, 12:\penalty0 2825--2830,
  2011.

\bibitem[Persson(2012)]{persson12a}
R.~A.~X. Persson.
\newblock {Note: Modification of the Gay-Berne potential for improved accuracy
  and speed}.
\newblock \emph{The Journal of Chemical Physics}, 136\penalty0 (22):\penalty0
  226101, 06 2012.
\newblock ISSN 0021-9606.
\newblock \doi{10.1063/1.4729745}.
\newblock URL \url{https://doi.org/10.1063/1.4729745}.

\bibitem[Schnitzer et~al.(1990)Schnitzer, Block, Berg, and
  Purcell]{schnitzer90a}
M.~Schnitzer, S.~Block, H.~Berg, and E.~Purcell.
\newblock Biology of the chemotactic response (armitage, jp \& lackie, jm eds)
  15--34, 1990.

\bibitem[Schulman et~al.(2017)Schulman, Wolski, Dhariwal, Radford, and
  Klimov]{schulman17a}
J.~Schulman, F.~Wolski, P.~Dhariwal, A.~Radford, and O.~Klimov.
\newblock Proximal policy optimization algorithms, 2017.

\bibitem[Sutton and Barto(2018)]{sutton98a}
R.~S. Sutton and A.~G. Barto.
\newblock \emph{Reinforcement Learning: An Introduction}.
\newblock The MIT Press, second edition, 2018.
\newblock URL \url{http://incompleteideas.net/book/the-book-2nd.html}.

\bibitem[Sutton et~al.(1999)Sutton, McAllester, Singh, and Mansour]{sutton99a}
R.~S. Sutton, D.~McAllester, S.~Singh, and Y.~Mansour.
\newblock Policy gradient methods for reinforcement learning with function
  approximation.
\newblock In S.~Solla, T.~Leen, and K.~M\"{u}ller, editors, \emph{Advances in
  Neural Information Processing Systems}, volume~12. MIT Press, 1999.

\bibitem[Tovey et~al.(2023{\natexlab{a}})Tovey, Lohrmann, Zimmer, Koppenhoefer,
  and Merkt]{swarmrl}
S.~Tovey, C.~Lohrmann, D.~Zimmer, S.~Koppenhoefer, and T.~Merkt.
\newblock {SwarmRL}, 2023{\natexlab{a}}.
\newblock URL \url{https://github.com/SwarmRL/SwarmRL}.

\bibitem[Tovey et~al.(2023{\natexlab{b}})Tovey, Zimmer, Lohrmann, Merkt,
  Koppenhoefer, Heuthe, Bechinger, and Holm]{tovey23c}
S.~Tovey, D.~Zimmer, C.~Lohrmann, T.~Merkt, S.~Koppenhoefer, V.-L. Heuthe,
  C.~Bechinger, and C.~Holm.
\newblock Environmental effects on emergent strategy in micro-scale multi-agent
  reinforcement learning, 2023{\natexlab{b}}.

\bibitem[Turner et~al.(2000)Turner, Ryu, and Berg]{turner00a}
L.~Turner, W.~S. Ryu, and H.~C. Berg.
\newblock Real-time imaging of fluorescent flagellar filaments.
\newblock \emph{Journal of bacteriology}, 182\penalty0 (10):\penalty0
  2793--2801, 2000.

\bibitem[van~der Maaten and Hinton(2008)]{maaten08a}
L.~van~der Maaten and G.~Hinton.
\newblock Visualizing data using t-sne.
\newblock \emph{Journal of Machine Learning Research}, 9\penalty0
  (86):\penalty0 2579--2605, 2008.
\newblock URL \url{http://jmlr.org/papers/v9/vandermaaten08a.html}.

\bibitem[Wadhams and Armitage(2004)]{wadhams04a}
G.~H. Wadhams and J.~P. Armitage.
\newblock Making sense of it all: bacterial chemotaxis.
\newblock \emph{Nature Reviews Molecular Cell Biology}, 5\penalty0
  (12):\penalty0 1024--1037, Dec 2004.
\newblock ISSN 1471-0080.
\newblock \doi{10.1038/nrm1524}.
\newblock URL \url{https://doi.org/10.1038/nrm1524}.

\bibitem[Watari and Larson(2010)]{watari10a}
N.~Watari and R.~G. Larson.
\newblock The hydrodynamics of a run-and-tumble bacterium propelled by
  polymorphic helical flagella.
\newblock \emph{Biophys J}, 98\penalty0 (1):\penalty0 12--17, Jan. 2010.

\bibitem[Weeks et~al.(1971)Weeks, Chandler, and Andersen]{weeks71a}
J.~D. Weeks, D.~Chandler, and H.~C. Andersen.
\newblock Role of repulsive forces in determining the equilibrium structure of
  simple liquids.
\newblock \emph{The Journal of chemical physics}, 54\penalty0 (12):\penalty0
  5237--5247, 1971.

\bibitem[Weik et~al.(2019)Weik, Weeber, Szuttor, Breitsprecher, de~Graaf,
  Kuron, Landsgesell, Menke, Sean, and Holm]{weik19a}
F.~Weik, R.~Weeber, K.~Szuttor, K.~Breitsprecher, J.~de~Graaf, M.~Kuron,
  J.~Landsgesell, H.~Menke, D.~Sean, and C.~Holm.
\newblock Espresso 4.0 -- an extensible software package for simulating soft
  matter systems.
\newblock \emph{The European Physical Journal Special Topics}, 227\penalty0
  (14):\penalty0 1789--1816, Mar 2019.
\newblock ISSN 1951-6401.
\newblock \doi{10.1140/epjst/e2019-800186-9}.
\newblock URL \url{https://doi.org/10.1140/epjst/e2019-800186-9}.

\end{thebibliography}

\appendix
\subsection{Anisotropic Friction Coefficients}
\label{subsec:anisotropic-friction}
In the case of the anisotropic agents, the translational friction coefficients are more challenging.
Consider a spheroidal agent with axial semiaxis, $r_{ax}$ and equatorial semiaxis, $r_{eq}$.
Let
\begin{equation}
    a = \max(r_{ax}, r_{eq}), \quad
    e = \sqrt{1 - \left(\frac{r_{eq}}{r_{ax}}^{2}\right)}, \quad
    L = \log\left(\frac{1 + e}{1 - e} \right).
\end{equation}
The translational friction can now be defined following the approach taken by~\citet{datta99a} where for a prolate particle
\begin{equation}
    \gamma^{ax}_{t} = 16\pi\mu a e^{3}\left[\left(1 + e^{2}\right)L - 2e  \right]^{-1}
\end{equation}
is the axial friction coefficient and
\begin{equation}
    \gamma^{eq}_{t} = 32\pi\mu a e ^{3}\left[2e + (3e^{2} - 1)L \right]^{-1}
\end{equation}
the equatorial, and for an oblate particle
\begin{equation}
    \gamma^{ax}_{t} = 8\pi\mu a e^{3} \left[e(1 - e^{2})^{\frac{1}{2}} - (1 - 2e^{2})\sin^{-1} e\right]^{-1}
\end{equation}
\begin{equation}
    \gamma^{eq}_{t} = 16\pi\mu a e^{3}\left[(1 + 2e^{2})\sin^{-1} e - e(1 - e^{2})^{\frac{1}{2}} \right]^{-1}
\end{equation}
are the axial and equatorial friction coefficients respectively.
Rotational friction factors were compute using the Perrin factors~\cite{koenig75a}.
To do so, we introduce further the aspect ratio
\begin{equation}
    p = \frac{r_{ax}}{r_{eq}},
\end{equation}
and
\begin{equation}
    \xi = \frac{\sqrt{|p^{2} - 1|}}{p}.
\end{equation}
With these two equations, we can derive the Perrin $S$ factors for prolate
\begin{equation}
S^{\text{prolate}} = 2\frac{\tanh^{-1} \xi}{\xi}
\end{equation}
and oblate
\begin{equation}
    S^{\text{oblate}} = 2\frac{\tan^{-1} \xi}{\xi}
\end{equation}
particles respectively.
Finally, we define the generic rotational friction coefficient for an equivalent sphere
\begin{equation}
    \gamma_{\text{sphere}} = 8\pi\mu r_{ax} r_{eq}^{2}.
\end{equation}
With these definitions, the equatorial and axial friction coefficients for both prolate and oblate particles can be derived as
\begin{equation}
    \prescript{}{\text{prolate/oblate}}\gamma^{eq}_{r} = \frac{4}{3}\frac{\frac{1}{p^{2}} - p^{2}}{2 - S^{\text{prolate/oblate}} \left[2 - \left(\frac{1}{p}\right)^{2} \right]}
\end{equation}
and
\begin{equation}
    \prescript{}{\text{prolate/oblate}}\gamma^{ax}_{r} = \frac{4}{3}\frac{p^{2} - 1}{2p^{2} - S^{\text{prolate/oblate}}}.
\end{equation}
For anisotropic particles, the translational friction coefficient in Equation~\eqref{eq:brownian_pos} becomes a tensorial quantity with $\gamma_{t} = \text{diag}(\gamma_t^{eq}, \gamma_t^{ax})$ in the comoving frame of reference of each particle in which the second coordinate points along the director $\vb{e}(t)$.
Since particles are fixed to two dimensions, rotation happens only around an equatorial axis such that for anisotropic particles $\gamma_r = \gamma_r^{eq}$ in \cref{eq:brownian_angle}.
The ESPResSo~\citep{weik19a} simulation package is used to numerically solve Equations~\eqref{eq:brownian_pos} and~\eqref{eq:brownian_angle} with a time-step $\delta t = \SI{0.005}{\second}$, the actions that determine $F(t)$ and $m(t)$ are updated every time slice $\Delta t = \SI{0.1}{\second}$.
In all cases, unless otherwise specified, when referring to time in this investigation, we refer to the time slice, i.e., the number of times an action is computed for each agent in the simulation.

\subsection{Anisotropic Interaction Potential}
\label{subsec:anisotropic-interaction}
For the prolate and oblate agents, we use the modified Gay-Berne potential for anisotropic particles~\cite{persson12a, gay81a}
\begin{equation}
    \centering
    V\left(\vb{u}_{i}, \vb{u}_{j}, \vb{r}_{ij}\right) = 
    \begin{cases}
     \epsilon\left(\vb{u}_{i}, \vb{u}_{j}, \vb{r}_{ij}\right) \left[ \left( \frac{\sigma_{0}}{r_{ij} - \sigma\left(\vb{u}_{i}, \vb{u}_{j}, \vb{r}_{ij}\right) + \sigma_{0}}\right)^{12} - \left( \frac{\sigma_{0}}{r_{ij} - \sigma\left(\vb{u}_{i}, \vb{u}_{j}, \vb{r}_{ij}\right) + \sigma_{0}}\right)^{6}\right], \quad &r_{ij}<4\sigma\max(l, l^{-1}),\\
    0, &\text{else},
    \end{cases}
\end{equation}
where $\vb{r}_{ij}$ is the vector distance between the center of mass of each particle, $\vb{u}_{i}$ is the direction vector of the $i^{\text{th}}$ particle, and $\epsilon\left(\vb{u}_{i}, \vb{u}_{j}, \vb{r}_{ij}\right)$ and $\sigma\left(\vb{u}_{i}, \vb{u}_{j}, \vb{r}_{ij}\right)$ are additional functions depending on the orientations of the particles.
We use the augmented forms of these function where
\begin{equation}
    \centering
    \epsilon\left(\vb{u}_{i}, \vb{u}_{j}, \vb{r}_{ij}\right) = \epsilon_{0}\left[1 +  \frac{d^{-1} - 1}{2} \left(||\vb{r}_{ij}\cdot\vb{u}_{1}||_{2} +  ||\vb{r}_{ij}\cdot\vb{u}_{2}||_{2}\right)\right]
\end{equation}
and
\begin{equation}
    \centering
    \sigma\left(\vb{u}_{i}, \vb{u}_{j}, \vb{r}_{ij}\right) = \sigma_{0}\left[1 +  \frac{l - 1}{2} \left(||\vb{r}_{ij}\cdot\vb{u}_{1}||_{2} +  ||\vb{r}_{ij}\cdot\vb{u}_{2}||_{2}\right)\right],
\end{equation}
where $d$ is ratio between the side-by-side binding energy and the end-to-end binding energy and $l$ is the aspect ration, which, in this work, was set to $3$ for the prolate particles and $\frac{1}{3}$ for the oblates.
We choose $\sigma_0 = \sigma$ and $\epsilon_0 =V_0$.

\subsection{Policy Examples}
\label{subsec:policy-examples}
Here we show example policy diagrams for each policy discussed in the manuscript.
\begin{figure}[ht]
    \centering
    \includegraphics[width=1.1\linewidth]{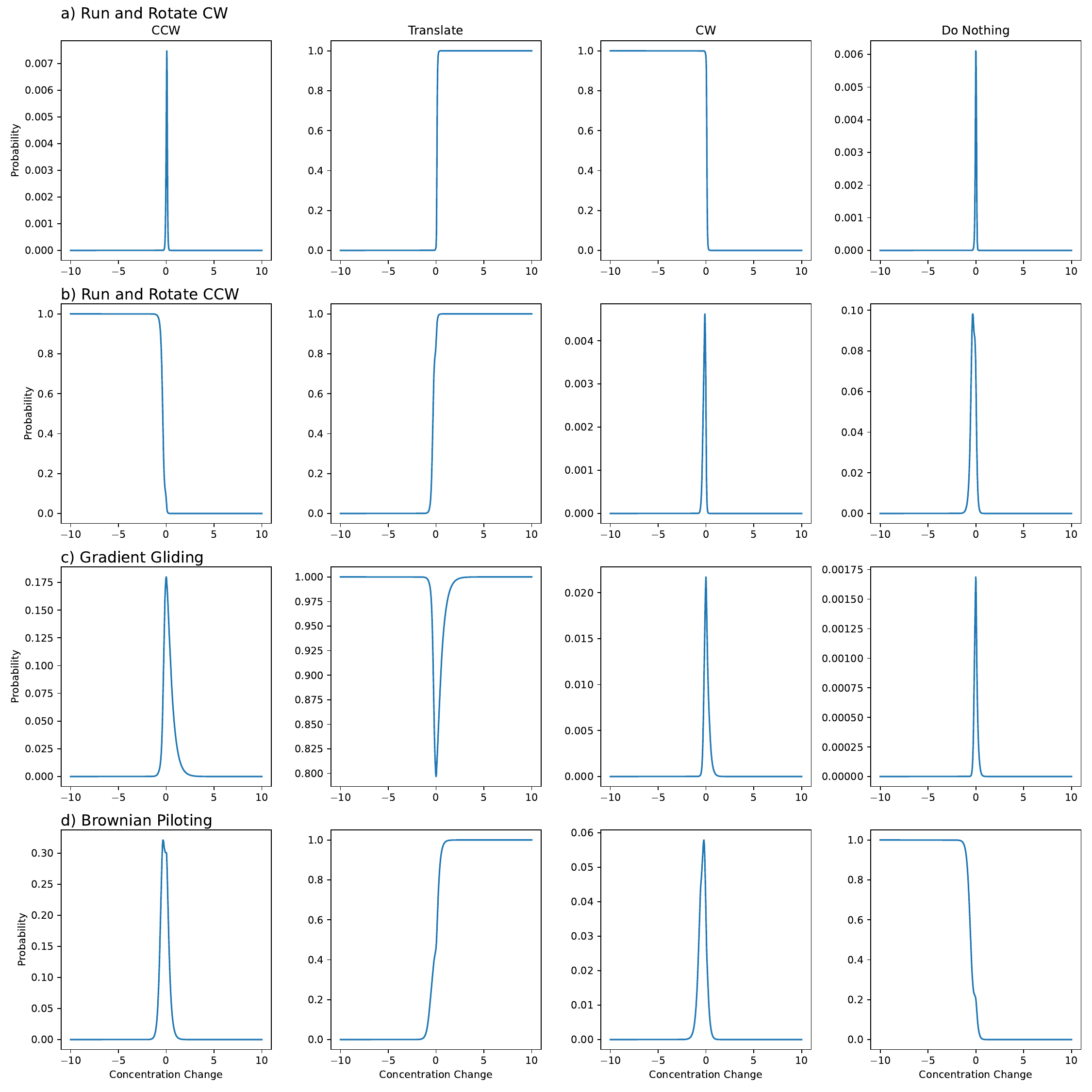}
    \caption{Examples of each emergent policy found during the investigation. a) and b) are the run and rotate policy for CW and CCW directions. You can see that as the input to the network becomes negative, the agents decide to rotate and as they are moving towards the source, they translate. c) The Just run, rarely rotate policy where for most inputs, the agent will translate, but far away from the source, when the input is small, the agent may also decide to rotate. d) Do nothing when negative and run when positive.}
    \label{fig:small-policy-examples}
\end{figure}

\subsection{Shape Studies}
\label{subsec:shape-studies}
As discussed in the manuscript, we performed the chemotaxis study for spherical, prolate, and oblate particles.
Here we display and discuss the results not presented in the main manuscript.
\subsubsection{Sphere}
While most of the sphere results are presented in the main manuscript, the raw policy plots are presented here.
Note, bl in the figure stands for body lengths.
\begin{figure}[ht]
    \centering
    \includegraphics[width=0.9\linewidth]{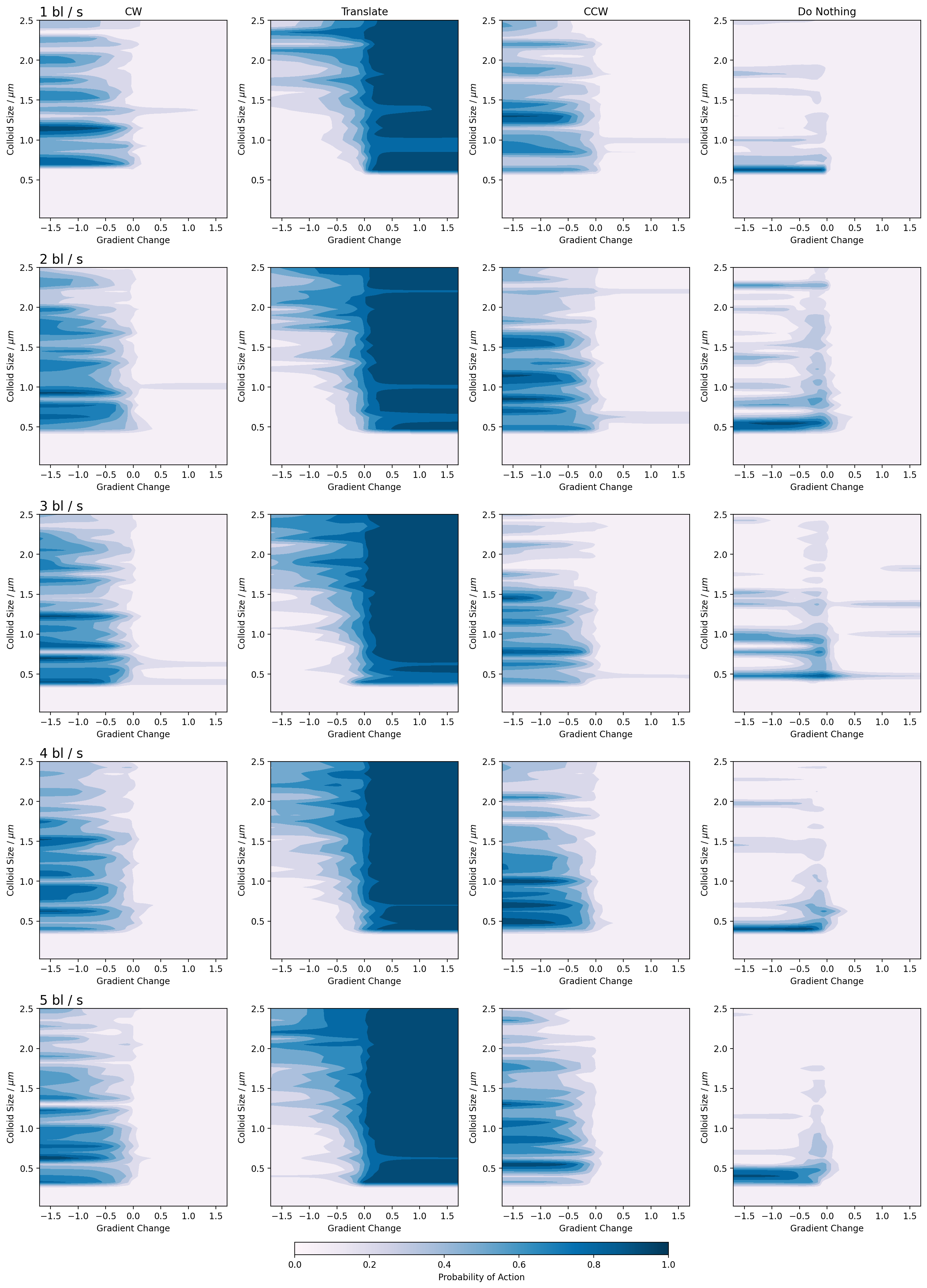}
    \caption{Emergent policy of the spherical microswimmers for all speeds and sizes. We see the development of several strategies depending on body size, notably, a run-and-tumble type strategy where colloids will either rotate or translate depending on a decrease or increase in concentration, respectively. Interestingly, once a colloid has learned to rotate either CW or CCW, it does not change this direction at any point during the runs.}
    \label{fig:policy-plots}
\end{figure}
The sphere policy diagrams outline the majority of the policies discussed in the main text.
On the x axis, the change in gradient is plotted and on the y, the colloid shape.
The colour of the diagram represents the probability of an action being taken and each column corresponds to a single action.
The rows are the different swim speeds descending from one to five.
The diagrams show the forbidden region in the chemotaxis below approximately $0.5 \mu m$.
After this point, we see the emergence of non-zero probabilities as the networks have learned to perform chemotaxis.
\subsubsection{Oblate}
The oblate particles demonstrated similar behaviour to the spherical colloids, not showing any unique policy deviations. 
\begin{figure}[ht]
    \centering
    \includegraphics[width=\linewidth]{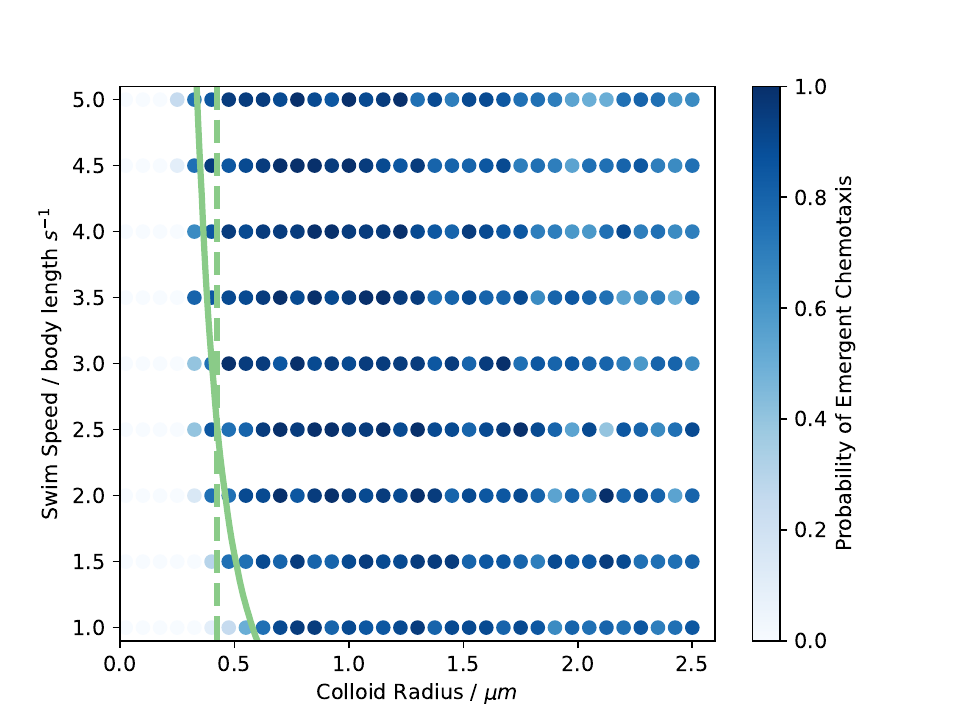}
    \caption{The probability of emergent chemotaxis for the oblate particles along with the theoretical boundaries from the P\'{e}clet numbers.}
    \label{fig:oblate-p-of-c}
\end{figure}
\begin{figure}[ht]
    \centering
    \includegraphics[width=0.8\linewidth]{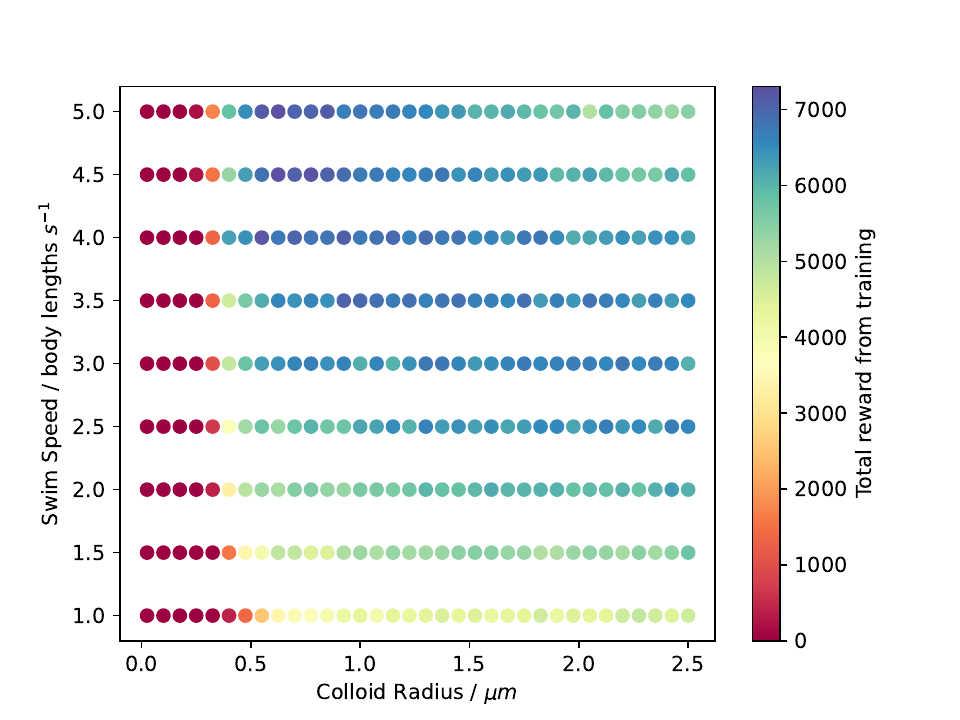}
    \caption{Reward phase diagram computer from the oblate simulations. In this case, we see the best training emerges again in the smaller but fast region of the diagram. While the width of the colloids is corrected for, more complex geometric conditions may be impacting these results.}
    \label{fig:oblate-reward}
\end{figure}

\begin{figure}[ht]
    \centering
    \includegraphics[width=0.8\linewidth]{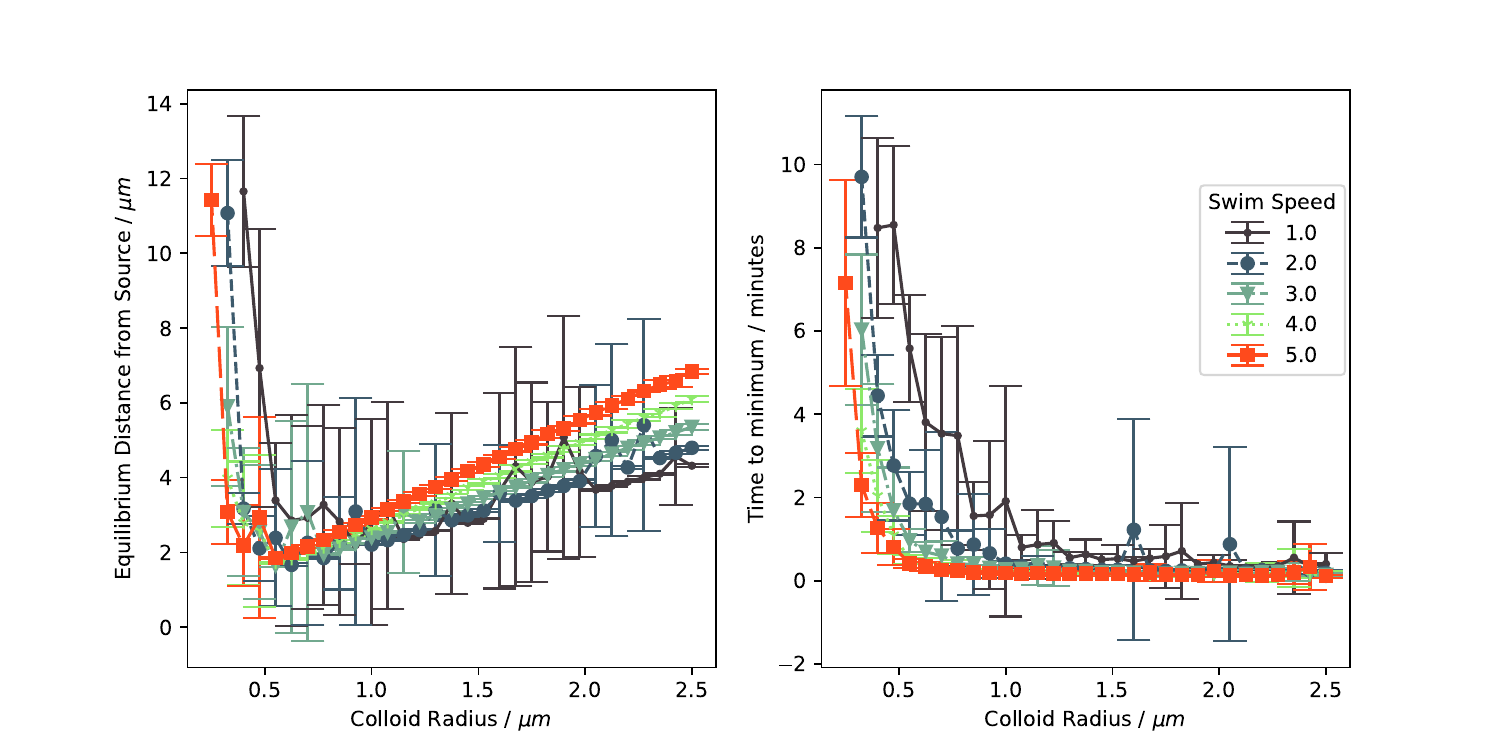}
    \caption{Policy efficacy for the oblate particles appears almost identical to the spherical particles.}
    \label{fig:oblate-effic}
\end{figure}

\begin{figure}[ht]
    \centering
    \includegraphics[width=0.9\linewidth]{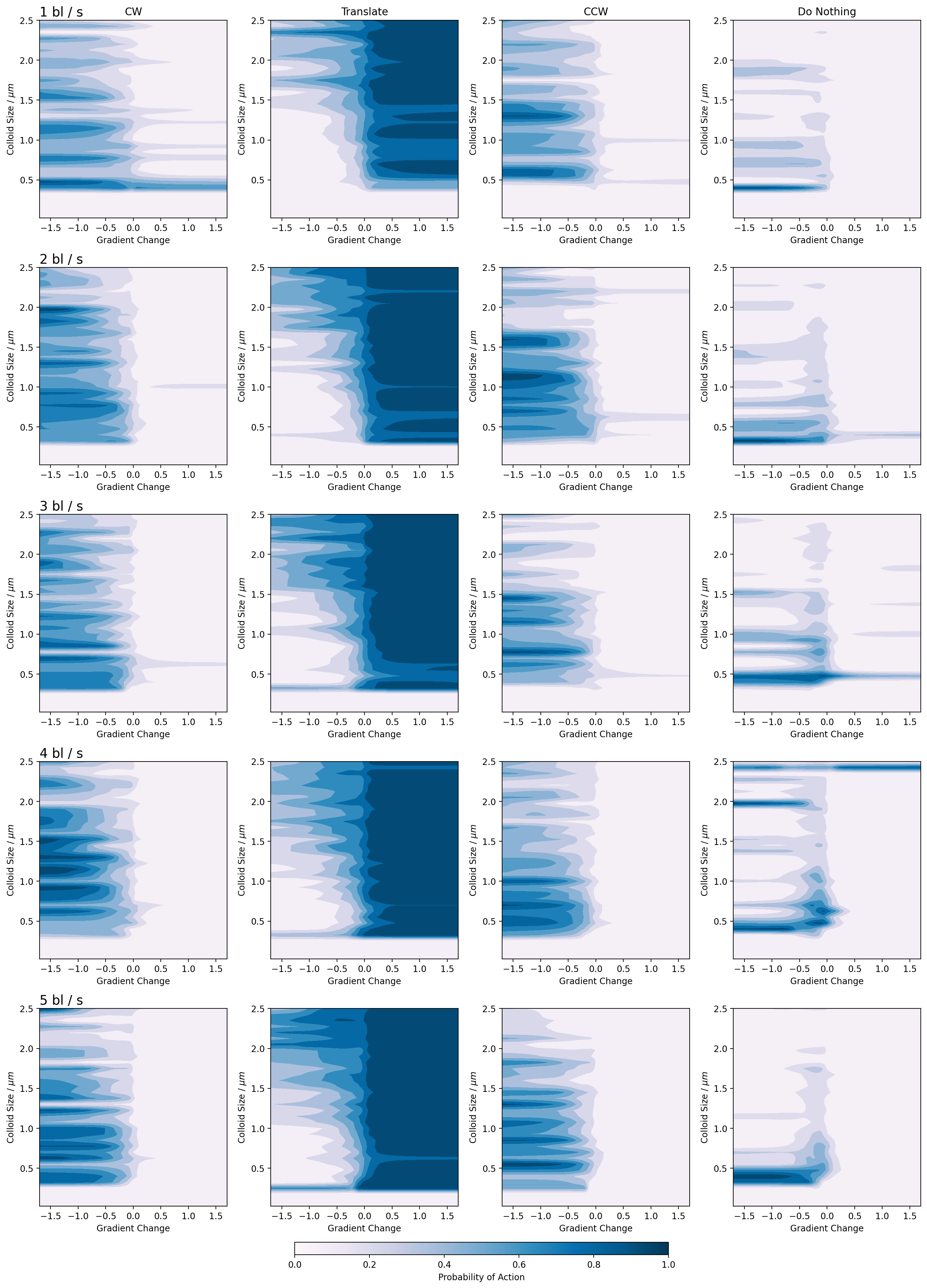}
    \caption{Emergent policy diagram for the oblate particles.}
    \label{fig:oblate-emerge}
\end{figure}

\subsubsection{Prolate}
The prolate simulations were similar to both the spherical and oblate studies with the exception of one very unique policy.
Mentioned in the main manuscript, we found an agent that appeared to perform no actions until the input to the network was small and positive.
Upon receiving such an input, the agent would translate.
More interestingly, this occured for a particle of size around $0.3 \mu m$, far below the theoretical boundaries for random forces to begin dominating the motion.
This was also the smallest agent in all simulations capable of achieving chemotaxis.
\begin{figure}[ht]
    \centering
    \includegraphics[width=\linewidth]{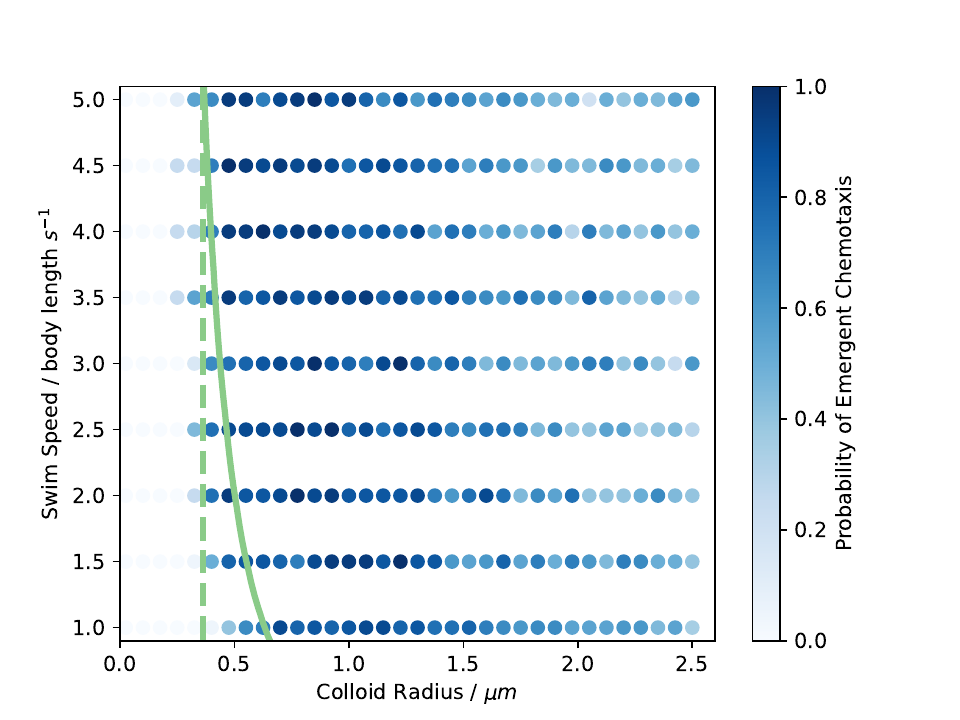}
    \caption{The probability of emergent chemotaxis for the prolate particles along with the theoretical boundaries from the P\'{e}clet numbers.}
    \label{fig:prolate-p-of-c}
\end{figure}

\begin{figure}[ht]
    \centering
    \includegraphics[width=0.8\linewidth]{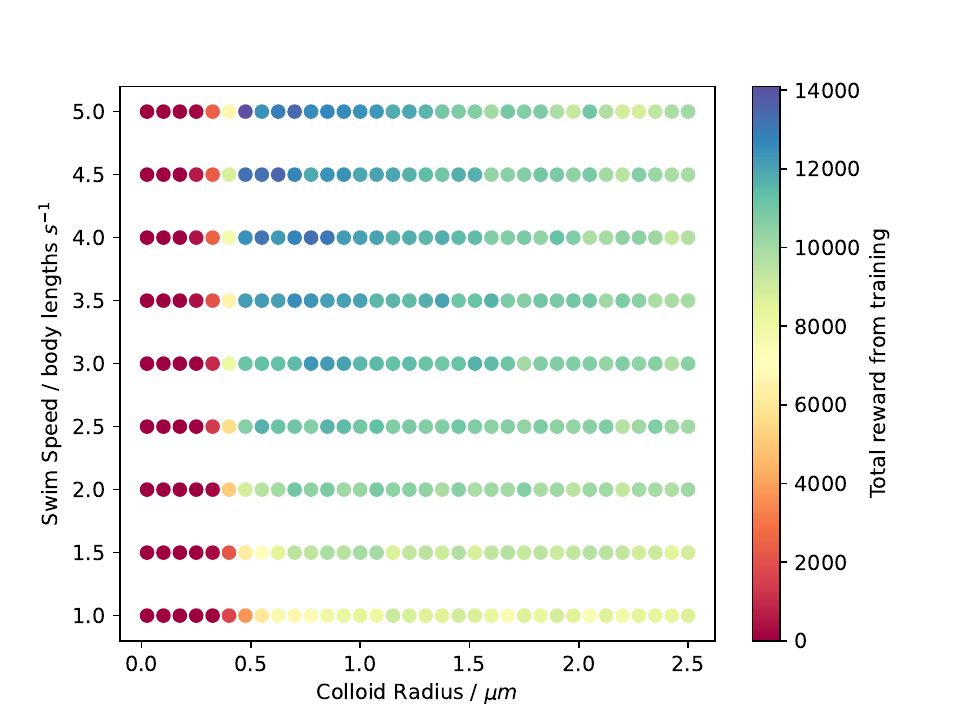}
    \caption{Reward phase diagram computer from the prolate simulations. In this case, we see the best training emerges again in the smaller but fast region of the diagram. While the width of the colloids is corrected for, more complex geometric conditions may be impacting these results.}
    \label{fig:prolate-rew}
\end{figure}

\begin{figure}[ht]
    \centering
    \includegraphics[width=0.8\linewidth]{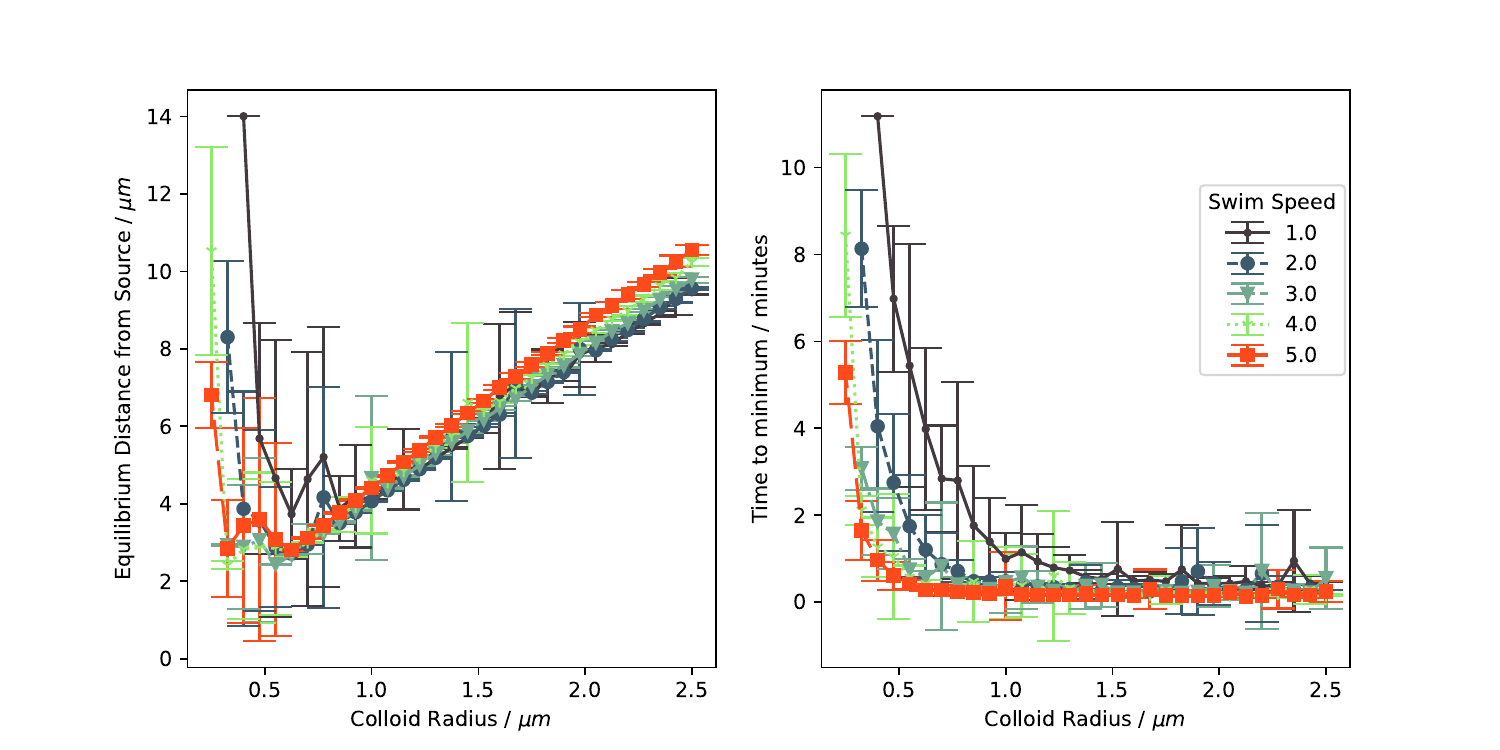}
    \caption{Policy efficacy for the prolate particles appears almost identical to the spherical particles.}
    \label{fig:prolate-eff}
\end{figure}

\begin{figure}[ht]
    \centering
    \includegraphics[width=0.9\linewidth]{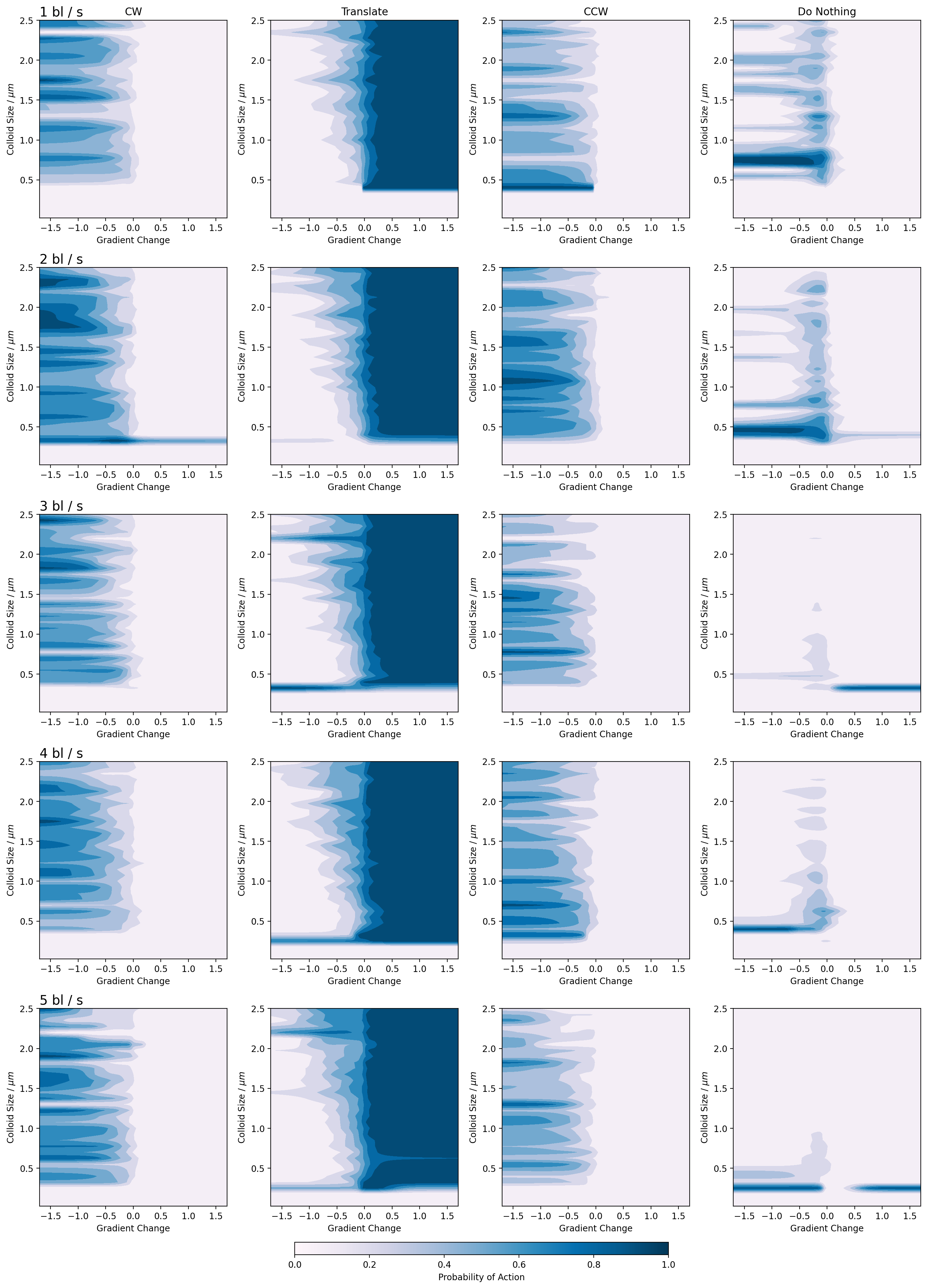}
    \caption{Emergent policy diagram for the prolate particles.}
    \label{fig:prolate-po}
\end{figure}

\end{document}